\documentclass{IEEEtran} 
\usepackage{bm, amssymb, subfigure, graphicx, amsthm, color}
\usepackage[cmex10]{amsmath}

\usepackage{footnote}
\usepackage{perpage}
\makeatletter
\def\blfootnote{\xdef\@thefnmark{}\@footnotetext}
\makeatother

\usepackage{breqn}
\newcommand{\mc}{\condition*}

\newtheorem{theorem}{Theorem}

\newtheorem*{corollary}{Corollary}

\renewcommand{\max}{\text{max}}
\renewcommand{\min}{\text{min}}
\newcommand{\btheta}{\bm{\theta}}

\newcommand{\bomega}{\bm{\omega}}
\newcommand{\bT}{\mathbf{T}}
\newcommand{\bq}{\mathbf{q}}
\newcommand{\by}{\mathbf{y}}
\newcommand{\bA}{\bm{A}}
\newcommand{\bx}{\mathbf{x}}
\newcommand{\ba}{\mathbf{a}}
\newcommand{\bs}{\mathbf{s}}
\newcommand{\bepsilon}{\bm{\epsilon}}
\newcommand{\bz}{\mathbf{z}}
\newcommand{\bdelta}{\bm{\delta}}
\newcommand{\bphi}{\bm{\phi}}
\newcommand{\bB}{\bm{\Phi}}
\newcommand{\bg}{\mathbf{g}}
\newcommand{\bX}{\mathbf{X}}

\newcommand{\bK}{\mathbf{K}}
\newcommand{\bomegatilde}{\tilde{\bm{\omega}}}
\newcommand{\bgtilde}{\tilde{\mathbf{g}}}
\newcommand{\be}{\mathbf{e}}
\newcommand{\bh}{\mathbf{h}}
\newcommand{\bv}{\mathbf{v}}
\newcommand{\bw}{\mathbf{w}}
\newcommand{\negspace}{\vspace{-0.075in}}

\begin{document}
\title{Compressive parameter estimation in AWGN} 

\author{Dinesh Ramasamy, Sriram Venkateswaran and Upamanyu Madhow}

\maketitle
\blfootnote{D. Ramasamy and U. Madhow are with the Department of ECE, University of California Santa Barbara. S. Venkateswaran is with Broadcom Corporation. Email:~\{dineshr,~sriram,~madhow\}@ece.ucsb.edu. This work was supported by the National Science Foundation through the grant CNS-0832154, by the Institute for Collaborative Biotechnologies through the grant W911NF-09-0001 from the U.S. Army Research Office and by the Systems on Nanoscale Information fabriCs (SONIC), one of six centers supported by the STARnet phase of the Focus Center Research Program (FCRP), a Semiconductor Research Corporation program sponsored by MARCO and DARPA.  The content of the information does not necessarily reflect the position or the policy of the Government, and no official endorsement should be inferred.}

\begin{abstract}

Compressed sensing is by now well-established as an effective tool for extracting sparsely distributed information, where sparsity is a {\it discrete} concept, referring to the number of dominant nonzero signal components in some basis for the signal space. In this paper, we establish a framework for estimation of {\it continuous-valued} parameters based on compressive measurements on a signal corrupted by additive white Gaussian noise (AWGN). While standard compressed sensing based on naive discretization has been shown to suffer from performance loss due to basis mismatch, we demonstrate that this is not an inherent property of compressive measurements. Our contributions are summarized as follows: (a) We identify the isometries required to preserve fundamental estimation-theoretic quantities such as the Ziv-Zakai bound (ZZB) and the Cram\'er-Rao bound (CRB). Under such isometries, compressive projections can be interpreted simply as a reduction in ``effective SNR." (b) We show that the threshold behavior of the ZZB provides a criterion for determining the minimum number of measurements for ``accurate" parameter estimation. (c) We provide detailed computations of the number of measurements needed for the isometries in (a) to hold for the problem of frequency estimation in a mixture of sinusoids. We show via simulations that the design criterion in (b) is accurate for estimating the frequency of a single sinusoid.

\end{abstract}

\section{Introduction}
\label{sec:intro}
Compressed sensing has proven remarkably successful in exploiting sparsity to extract information from signals with only a small number of measurements. The standard approach has two stages. First, take multiple random projections of the signal, with the number of projections growing linearly with the sparsity and only logarithmically with the dimensionality of the signal. Then, use one among a variety of recovery algorithms, such as $\ell_1$ reconstruction/Orthogonal Matching Pursuit (OMP), to estimate the signal from the random projections. In this standard framework, sparsity is an inherently {\it discrete} concept: the number of nonzero signal components in some basis has to be small compared to the dimension of the signal. In this paper, we investigate the effectiveness of compressive measurements in estimating {\it continuous valued} parameters from signals that are corrupted by AWGN, when the dimensionality of the parameter set is much smaller than the signal dimension.

It is possible to apply standard compressed sensing to continuous-valued parameter estimation, but it does not perform well. Consider the fundamental problem of estimating the frequencies in a mixture of sinusoids. Typically, the number of sinusoids is much smaller than the number of samples and, therefore, the signal is sparse in the frequency domain. However, the conventional compressed sensing framework does not apply directly, since it requires the signal to be sparse over a {\it finite} basis, whereas the frequencies could lie anywhere on a continuum.  Straightforward application of compressed sensing recovery algorithms after discretizing the set of frequencies has been shown to result in error floors due to ``basis mismatch'' and the consequent spectral leakage \cite{yuejie_chi_sensitivity_Basis_Mismatch_2011}. This observation raises some fundamental questions. Do compressive measurements preserve all the information needed for continuous valued parameter estimation? If so, under what conditions? How many measurements do we require to satisfy these conditions? In this paper, we establish a systematic framework that addresses these questions for parameter estimation based on signals corrupted by AWGN.

\noindent {\bf Contributions:} We first identify fundamental structural properties for compressive estimation in AWGN, and then illustrate them by explicit computation for frequency estimation for a mixture of sinusoids.\\
{\it Isometries for estimation:} Suppose we want to estimate a $K$ dimensional parameter $\btheta = (\theta_{1},\theta_{2},\ldots,\theta_{K})$ from projections of an $N$ dimensional signal $\bx(\btheta)$ in AWGN. When we make all $N$ measurements, fundamental bounds on the estimation error variance, such as the Ziv-Zakai bound (ZZB) and the Cram\'er Rao bound (CRB), relate the geometry of the signal manifold to the best achievable performance. From the ZZB, we can infer that ``coarse'' estimation depends on the pairwise distances $\Vert \bx(\btheta) - \bx(\btheta') \Vert~\forall \btheta,\btheta'$, while the CRB tells us that ``fine'' estimation depends on norms of linear combinations of the partial derivatives $\{\partial \bx/\partial \theta_{k}\}$ (vectors in the tangent plane, which are the limit of differences as $\btheta \rightarrow \btheta'$). We extend these observations to compressive estimation by replacing the signal manifold $\bx(\btheta)$ by $\bA \bx(\btheta)$, where $\bA$ is the compressive measurement matrix containing the random projection weights. We identify the isometries required to ensure that the geometry (and hence the structure of the ZZB and CRB) is roughly unaltered after the compressive projection.  We also note that, if these isometries hold, then the only consequence of compressive projection onto a subspace of dimension $M$ is an SNR penalty of $M/N$.  This is because each random projection captures $1/N$ of the signal energy on average (normalizing such that the noise variance is unchanged). Specifically, we show that if the measurement matrix $\bA$ satisfies the pairwise isometry property (PIP) ($\Vert \bA \bx(\btheta) - \bA \bx(\btheta') \Vert \approx \sqrt{M/N} \Vert \bx(\btheta) - \bx(\btheta') \Vert$), the ZZB with compressive measurements is approximately equal to the ZZB with all $N$ measurements, except for the SNR penalty of $M/N$. We prove an analogous result for the CRB when $\bA$ guarantees tangent plane isometry ($\Vert \bA \sum_{k} a_{k} \partial \bx(\btheta)/\partial \theta_{k} \Vert \approx \sqrt{M/N} \Vert \sum_{k} a_{k} \partial \bx(\btheta)/\partial \theta_{k} \Vert, ~\forall {a_{k}}$), which is a weaker requirement than pairwise isometry. \\
{\it Number of measurements:} When the preceding isometries hold, we can use their relationship to the ZZB/CRB to obtain a tight prediction on the number of measurements necessary for successful compressive estimation. It is known that nonlinear estimation problems exhibit a threshold behavior with the SNR which is closely mirrored by the threshold behavior of the ZZB. We employ this observation to predict the number of measurements required to avoid performance floors, since the the effective SNR with compressive measurements increases linearly with the number of measurements.\\
{\it Computations for sinusoidal mixtures:} While the preceding results reveal the structure of compressive estimation, computation of the number of measurements required to achieve the desired isometries and to avoid performance floors requires a problem-specific analysis. To this end, we consider the fundamental problem of frequency estimation for a mixture of sinusoids.  For estimating $K$ frequencies from $N$ samples, we show that: (a) $O\left(K \log (NK \delta^{-1}) \right)$ measurements suffice to provide tangent plane isometries, where $\delta$ depends on the frequency separation between the sinusoids in the mixture ($\delta$ vanishes when any two of the $K$ frequencies approach one another). (b) $O\left( K \log (NK \delta^{-1}) \right)$ measurements suffice to provide pairwise isometries between two sets of frequencies $\bomega = (\omega_{1},\omega_{2},\ldots,\omega_{K})$ and $\bomega^{\prime} = (\omega_{1}^{\prime},\omega_{2}^{\prime},\ldots,\omega_{K}^{\prime})$ that are ``well-separated.'' Here $\delta$ depends only on the frequency separation between the sinusoids in the mixture of $2K$ sinusoids $(\bomega,\bomega^{\prime})$, and vanishes when any two frequencies in $(\bomega,\bomega^{\prime})$ approach one another. Therefore, with $O\left( K \log (NK \delta^{-1}) \right)$ compressive measurements, we can preserve the ``well-separated" geometry of the frequency estimation problem. The tangent plane isometry results (a) indicate that when the $K$ frequencies in $\bomega$ themselves are ``well-separated", compressive measurements preserve the ``fine" geometry of the frequency estimation problem (and therefore the CRB). We strengthen these results for a single sinusoid ($K=1$), exploiting the continuity of the sinusoidal manifold to show that $O(\log N)$ measurements suffice to guarantee pairwise isometry between sinusoids at {\it any} two frequencies $\omega,\omega^{\prime}$ (by merging the ``well-separated" and ``fine" regimes). We also show that the criterion for prediction of the number of measurements, based on the threshold behavior of the ZZB, is tight, by evaluating the performance of an algorithm which closely approximates the MAP estimator. The algorithm works in two stages: first, from a discrete set of frequencies, we pick the one that fits the observations best and, then, we perform local refinements using Newton's method. 


\section{Related work}
\label{sec:relWork}

The goal of standard compressed sensing \cite{candes2,donoho} is to recover signals which are sparse over a finite basis with significantly fewer measurements than the dimension of the observation space.  Signal recovery requires that the measurement matrix must satisfy the Restricted Isometry Property (RIP): the distance between any two sparse signals must be roughly invariant under the action of the matrix. If the RIP is satisfied, sparse signals can be recovered efficiently using techniques such as Orthogonal Matching Pursuit (OMP) and $\ell_1$-norm minimization.  Reference \cite{baraniuk_simple_RIP_2008} used the Johnson-Lindenstrauss (JL) lemma to provide a simple proof  that $O(K \log N)$ random projections suffice to establish RIP for recovering $K$-sparse vectors in $\mathbb{R}^N$. We briefly summarize the key ideas, since we use an analogous approach in establishing pairwise isometry for the mixture of sinusoids example discussed in this paper. The JL lemma states that, to approximately preserve the pairwise distances between $P$ points after random projections (with the weights chosen from appropriate distributions, such as i.i.d. $\pm 1$ \cite{achlioptas_database-friendly_2001}), we need $O(\log P)$ such projections. However, to provide an RIP for compressive measurement matrices, the distances between any two $K$-sparse vectors must be preserved. Since the number of such vectors is infinite, the JL lemma cannot be applied directly. However, the desired RIP result is established in \cite{baraniuk_simple_RIP_2008} by discretizing the set of $K$-sparse vectors sufficiently finely, applying the JL lemma to the resulting discrete set of points, and then exploiting continuity to provide isometries for the remaining points.

For compressive estimation of continuous-valued parameters, sparsity corresponds to the dimension of the parameter space $K$ being significantly smaller than that of the observation space $N$. This problem was perhaps first investigated in \cite{baraniuk_random_2009}, which identifies that the analogue of the RIP here is the pairwise isometry property considered in the present paper. However, it does not relate this property to estimation-theoretic bounds as done here.
Reference \cite{baraniuk_random_2009} also shows that compressive measurements guarantee pairwise $\epsilon$-isometry for a signal manifold with probability $1-\rho$, as long as the number of measurements $M$ satisfies
\begin{equation}
M = O \left(\epsilon^{-2} \log(1/\rho) K \log\left(NVR \tau^{-1} \epsilon^{-1}\right)\right),
\label{eq:baraniukFormula}
\end{equation}
where $V,R,\tau$ are properties of the signal manifold ($1/\tau$ is the condition number which is a generalization of the radius of curvature, $R$ is the geodesic covering regularity and $V$ is the volume). However, to the best of our knowledge, it is difficult to specify how $\{\tau, V,R\}$ scale with the parameters $N$ and $K$ in general.  In this paper, therefore, we provide a self-contained derivation of the number of measurements required to preserve these isometries when the signal manifold consists of a mixture of sinusoids in Section \ref{sec:multipleSinusoids}.  Compressive parameter estimation has also been studied in \cite{wakin_manifold_estimation_2010}; however, since the noise model there is adversarial, the results are pessimistic for many practical applications in which a Gaussian model for the noise is a good fit.

Algorithms to estimate the frequencies in a mixture of sinusoids from compressive measurements are proposed and evaluated in \cite{Duarte_2012_SpectralCS,fannjiang:superresolution:compressive}. Both of these papers assume that the sinusoids have a minimum frequency separation and \cite{fannjiang:superresolution:compressive} further assumes that the frequencies come from an oversampled DFT grid. They propose variants of standard compressed sensing algorithms, such as Orthogonal Matching Pursuit (OMP) and Iterative Hard Thresholding (IHT), which rely on the sinusoids' frequencies not being too close. As mentioned earlier, restricting the frequency estimation to a discrete grid in this fashion results in performance floors, as studied in great detail in \cite{yuejie_chi_sensitivity_Basis_Mismatch_2011}. However, as shown in this paper and in our earlier conference papers \cite{asilomar:CRLB:frequency,allerton:tracking}, it is possible to avoid such performance floors, and to attain the CRB, by local refinements based on Newton-like algorithms after grid-based coarse estimation. A one-shot quadratic refinement is also proposed in \cite{Duarte_2012_SpectralCS} to improve estimates of off-grid frequencies. 

We characterize the structure of compressive estimation here in terms of that of the original problem.  However, in many cases, an estimation-theoretic understanding
of the original problem is incomplete: in particular, for the mixture of sinusoids model, a characterization of the difficulty
of the problem in terms of the minimum separation of frequencies in $\bomega$ remains an ongoing effort 
\cite{Recht_2013_with_Noise_no_CS,Recht_2012_denoising,2012_Recht_CS_Off_the_Grid}, as discussed in more detail below.

The problem of estimating frequencies in a mixture of sinusoids from noise-free compressive measurements is studied in \cite{2012_Recht_CS_Off_the_Grid}. While the frequencies can come from the $[0,2\pi)$ continuum, \cite{2012_Recht_CS_Off_the_Grid} requires that they are ``well-separated" (four times the DFT spacing of $2\pi/N$). When this condition is met, it is shown that atomic-norm denoising (cast as a semi-definite program) correctly estimates the frequencies in the mixture. The same $4\times(2\pi/N)$ frequency separation is shown to be necessary for recovering frequencies over a continuum with noisy measurements of all $N$ samples (not compressive) in \cite{Recht_2013_with_Noise_no_CS,Recht_2012_denoising}. It is interesting to note that even when all $N$ samples are observed, the same minimum frequency separation is necessary for stable recovery. This falls in line with the observations that we make on the equivalence (except for an SNR penalty) of the ``difficulty" in estimation using compressive measurements and uncompressed measurements (all $N$ samples) by relating corresponding estimation error bounds. 

To the best of our knowledge, other than our conference paper \cite{asilomar:CRLB:frequency}, this is the first paper to relate isometry conditions to estimation-theoretic bounds for compressive parameter estimation, and to show that, in the AWGN setting, the only effect of compressive measurements when appropriate isometry conditions are satisfied is an SNR penalty of $M/N$. The $M/N$ SNR penalty due to compressive measurements has also been noted in \cite{Eldar_Noise_folding_2011}, but we go further and make the connection between isometries and estimation bounds. Isometries and SNR loss for signal detection were
considered in \cite{nowak:CompressiveDetection}, but we believe that the present paper is the first to address these for the general problem of parameter estimation in AWGN. 

This paper goes beyond the results in our conference paper \cite{asilomar:CRLB:frequency} in multiple ways. First, we establish a connection between the pairwise isometry property and the Ziv-Zakai bound. We then show how the connections between the ZZB and CRB, together with the isometry conditions, can be used to predict the number of measurements required for accurate compressive estimation. We also characterize the number of measurements needed to provide isometry guarantees for a mixture of sinusoids unlike \cite{asilomar:CRLB:frequency}, which only deals with a single sinusoid. Finally, the isometry guarantees provided in \cite{asilomar:CRLB:frequency} for a pair of sinusoids require their frequencies to be ``well-separated''. Here, we close the gap and provide such an isometry for any pair of frequencies.  

In the algorithm description and numerical illustrations in this paper, we restrict attention to a single sinusoid in order to illustrate the fundamental features of compressive estimation.  However, as described in detail in our conference papers \cite{asilomar:CRLB:frequency,allerton:tracking}, our algorithmic approach (discrete grid followed by Newton refinement) extends easily to estimate the frequencies of multiple sinusoids. While the latter is a canonical problem of fundamental interest, it is worth noting that an important application that motivates us is the problem of adapting very large antenna arrays \cite{ITA2012, allerton:tracking}. Compressive parameter estimation in this context exploits the relatively small number of dominant multipath rays in order to estimate the spatial frequencies (and hence the angles of arrival) for these rays.

While we focus on compressive estimation based on a finite-dimensional signal, there has been significant research on the processing of continuous time signals exhibiting some measure of sparsity, sometimes termed ``finite rate of innovation'' (FRI) signals \cite{maravic_vetterli_FRI_2005}. Sampling strategies for parameter estimation for such signals are studied in \cite{Eldar_FRI_sampling_CRB_2012}, using the CRB as the performance metric. The benefits of such good sampling strategies coupled with compressive processing at the analog front end are investigated in \cite{eldar_2012Xampling}. The isometry conditions derived in the present paper could potentially provide a systematic framework for design of compressive analog front ends for FRI signals.

\vspace{0.075in}
\noindent \textbf{Outline:}  We begin in Section \ref{sec:modelIsometries} by stating the compressive parameter estimation problem in AWGN and the isometry properties needed for successful estimation. In Section \ref{sec:ParameterEstimationReview} we review bounds on parameter estimation in AWGN. The relationship between these estimation error bounds (CRB/ZZB) and the isometry properties are brought out in Section \ref{sec:isometryBoundRelation}. In Section \ref{sec:frequencyestimation:bounds}, we consider the problem of estimating the frequency of a sinusoid. We show how the threshold behavior of the ZZB can predict the number of compressive measurements needed to avoid error floors. Section \ref{sec:multipleSinusoids} derives the number of measurements needed to guarantee these isometry conditions for the problem of frequency estimation from a mixture of $K$ sinusoids and concludes by focussing on the single sinusoid case ($K=1$).
\section{Compressive measurements}
\label{sec:modelIsometries}

We begin by presenting the model for compressive measurements and providing the intuition behind two isometry conditions that are necessary for successful parameter estimation.

Consider the problem of estimating $\btheta\in\Theta\subseteq\mathbb{R}^K$ from noisy measurements of a differentiable manifold $\bx(\btheta)\in\mathbb{C}^N$. The conventional estimation problem involves measuring all $N$ elements of $\bx(\btheta)$ individually. In vector notation, the measurements are given by:
\begin{dmath}
\by = \bx(\btheta) + \bz \mc{\bz\sim\mathcal{CN}\left(\mathbf{0},\sigma^2\mathbb{I}_N\right)}.
\label{eq:normal:measurement:model}
\end{dmath}
In contrast, with compressive measurements, we only observe $M \ll N$ noisy projections of the manifold $\bx(\btheta)$. Therefore, we have
\begin{dmath}
\by = \bA\bx(\btheta) + \bz \mc{\bz\sim\mathcal{CN}\left(\mathbf{0},\sigma^2\mathbb{I}_M\right)},\label{eq:compressive:measurement:model}
\end{dmath}
where $\bA \in \mathbb{C}^{M\times N}$, which specifies the projection weights, is called the compressive measurement matrix. The elements of $\bA$ are chosen independently from zero-mean distributions of variance $1/N$ for which certain concentration results (we comment on this later) are available. Examples of such distributions include $\text{Uniform}\{\pm 1/\sqrt{N}\}$, Gaussian and $\text{Uniform}\{\pm 1/\sqrt{N},\pm j/\sqrt{N}\}$. When the matrix $\bm{A}$ satisfies certain isometry conditions, we can successfully estimate $\btheta$ from $M \ll N$ measurements. We first explain why these conditions are helpful intuitively and then define them formally.

The Maximum Likelihood (ML) estimator \cite{van2013detection} of $\btheta$ for the model in (\ref{eq:compressive:measurement:model}) is given by
\begin{align}
\hat{\btheta} & = \arg\underset{\btheta^{\prime}}{\min}\ \left\Vert\mathbf{y}- \bm{A}\mathbf{x}(\btheta^{\prime})\right\Vert\\
& =  \arg\underset{\btheta^{\prime}}{\min}\ \left\Vert \bm{A}\mathbf{x}(\btheta)- \bm{A}\mathbf{x}(\btheta^{\prime})+\mathbf{z}\right\Vert.\label{eq:ML:estimator:compressive:expand}
\end{align}
If the number of measurements is too small and $\bm{A}$ has a large nullspace, it is possible that $\Vert \bm{A} \left(\mathbf{x}(\btheta) - \mathbf{x}(\btheta') \right) \Vert \approx 0$ even when $\Vert \mathbf{x}(\btheta) - \mathbf{x}(\btheta') \Vert$ is large. Thus, with small amounts of noise $\mathbf{z}$, the optimizing parameter $\hat{\btheta}$ could be drastically different from the true parameter $\btheta$, resulting in large errors. This problem can be avoided if the matrix $\bm{A}$ preserves the geometry of the estimation problem by ensuring that the distance between $\mathbf{x}(\btheta)$ and $\mathbf{x}(\btheta')$ remains approximately unaltered under its action. Specifically, if we have,
\begin{dmath}
\Vert \bA \left(\bx(\btheta) - \bx(\btheta') \right) \Vert \propto \Vert \bx(\btheta) - \bx(\btheta') \Vert \mc{\forall \btheta, \btheta'\in\Theta,}\label{eq:approx:preserve:geom}
\end{dmath}
we see from (\ref{eq:ML:estimator:compressive:expand}) that the ML estimate at high SNR from $M$ compressive measurements roughly coincides with the estimate we would have obtained with (\ref{eq:normal:measurement:model}), where we have access to {\it all} $N$ measurements of $\mathbf{x}(\btheta)$. The {\it pairwise $\epsilon$-isometry property} captures this idea of distance preservation precisely.

\noindent \textbf{Pairwise $\epsilon$-isometry property:} The matrix $\bm{A}$ satisfies the pairwise $\epsilon$-isometry property ($\epsilon < 1$) for the signal model $\mathbf{x}(\btheta)$ if 
\begin{align}\label{eq:PIP}
	\sqrt{\frac{M}{N}}(1-\epsilon) \leq \frac{\left\Vert \bm{A}\mathbf{x}(\btheta_1) - \bm{A}\mathbf{x}(\btheta_2) \right\Vert}{\left\Vert \mathbf{x}(\btheta_1) - \mathbf{x}(\btheta_2) \right\Vert} \leq& 	\sqrt{\frac{M}{N}}(1+\epsilon)\nonumber\\& \forall \  \btheta_1, \btheta_2\in\Theta.
\end{align}
We now motivate the isometry constants $\sqrt{M/N}(1-\epsilon)$ and $\sqrt{M/N}(1+\epsilon)$. Let $\mathbf{w}_{i}^H$ denote the $i$-th row of $\bA$. Consider a single random projection of a signal $\bv$ onto the weights $\mathbf{w}_{i}$ that have been chosen independently from zero-mean distributions of variance $1/N$. The average energy in the projection is $1/N$ of the energy in the signal $\mathbf{v}$: $\mathbb{E}\left|\mathbf{w}_{i}^H\mathbf{v}\right|^2 = (1/N)\left \Vert\bv\right \Vert ^2$. Thus, $M$ compressive measurements capture $M/N$ of the signal energy on average: $\mathbb{E}\Vert \bA \bv \Vert^{2} = (M/N) \Vert \bv \Vert^{2}$. Thus, for compressive measurements, it is natural to define the pairwise isometry property with the constants $\sqrt{M/N}(1-\epsilon)$ and $\sqrt{M/N}(1+\epsilon)$.

When the elements of $\bA$ are drawn from appropriate distributions, for any particular realization of the measurement matrix $\bA$, $\Vert \bA \bv \Vert^{2}$ concentrates around its expected value $(M/N) \Vert \bv \Vert^{2}$ with high probability. Specifically, for any $\bv\in\mathbb{C}^N$: 
\begin{equation}
\Pr\left[\left|\frac{N}{M}\Vert \bA \bv \Vert^{2}- \Vert \bv \Vert^{2}\right|>\delta\right]<C\exp\left(-M~c(\delta)\right),\label{eq:concentration:projection}
\end{equation}
with constants $C$ and $c(\delta)$ that depend only on the distribution from which the elements of $\bA$ are picked from. For example, when the elements of $\bA$ are picked i.i.d from $\text{Uniform}\{\pm 1/\sqrt{N},\pm j/\sqrt{N}\}$ or $\text{Uniform}\{\pm 1/\sqrt{N}\}$ or $\mathcal{N}(0,1/N)$, we can show that $C=4$ and 
\begin{equation}
c(\delta) = \delta^2/4-\delta^3/6. \label{eq:concentration:rate}
\end{equation}
These concentration results are typically used to prove the pairwise isometry property (\ref{eq:PIP}). Refer \cite{vershynin_introduction_2010} for a class of distributions (this includes all sub-gaussian distributions) for which such results are available.
 
We note that a particular instance of a randomly generated measurement matrix need not satisfy the pairwise isometry property for the signal manifold $\bx(\btheta)$. However, when the number of measurements $M$ is sufficiently large, \cite{baraniuk_random_2009} shows that the pairwise $\epsilon$-isometry property can be satisfied with arbitrarily high probability (the proof involves the use of concentration results (\ref{eq:concentration:projection}) on carefully chosen samples on the manifold). 

A weaker notion of distance preservation is the {\it tangent plane isometry property} that is particularly useful when we wish to refine an estimate $\hat{\btheta}$ that is ``close" to the true parameter value.  In this case, since we are interested only in the ML cost surface around the true parameter $\btheta$, it suffices to preserve the geometry of the estimation problem in the vicinity of $\btheta$ by ensuring that the distances between $\mathbf{x}(\btheta')$ and $\mathbf{x}(\btheta)$ for $\btheta' \rightarrow \btheta$ are preserved under the action of $\bA$. This is captured by the tangent plane isometry property defined as follows. 

\noindent \textbf{Tangent plane $\epsilon$-isometry property:} The matrix $\bm{A}$ satisfies the tangent plane $\epsilon$-isometry property ($\epsilon < 1$) for the signal model $\mathbf{x}(\btheta)$ if 
\begin{align}
	\sqrt{\frac{M}{N}}(1-\epsilon)& \leq  \frac{\left\Vert\bm{A}\sum a_m(\partial \mathbf{x}(\btheta)/\partial \theta_m) \right\Vert}{\left\Vert \sum a_m(\partial \mathbf{x}(\btheta)/\partial \theta_m) \right\Vert} \leq \sqrt{\frac{M}{N}}(1+\epsilon)\nonumber\\ 
	&\forall\ [a_{1}, a_{2}, \ldots, a_{K}]^T\in\mathbb{R}^K\text{\textbackslash}\{\mathbf{0}\},\ \forall \btheta\in\Theta \label{eq:tangent_plane:isometry}
\end{align}

By letting $\btheta_{2} \rightarrow \btheta_{1}$ in the definition of the pairwise $\epsilon$-isometry property, we see that a matrix $\bm{A}$ which satisfies the pairwise isometry property for the signal model $\mathbf{x}(\btheta)$ also satisfies the tangent plane isometry, thereby confirming that tangent plane isometry is a weaker notion of distance preservation.

\section{Parameter estimation in AWGN}
\label{sec:ParameterEstimationReview}
We now review classical bounds on parameter estimation in AWGN that we relate to the isometry properties in the next section.

 Consider the problem of estimating a parameter $\btheta\in\Theta\subseteq\mathbb{R}^K$ from noisy observations of the differentiable manifold $\bs(\btheta)\in\mathbb{C}^M$. The observations are given by: 
\begin{equation}
\by = \bs(\btheta) + \bz,\ \bz\sim\mathcal{CN}(\mathbf{0},\sigma^2\mathbb{I}_M).
\label{eq:uncomp:meas:model}
\end{equation}
For this measurement model,  
\begin{equation}
p(\by|\btheta) =(\pi\sigma^2)^{-M} \exp\left(\left\Vert\by-\bs(\btheta)\right\Vert^2/\sigma^2\right).
\label{eq:uncomp:meas:model:density}
\end{equation}

For the observations $\by$, let $\hat{\btheta}(\by)$ be an estimate of $\btheta$. Given a weight vector $\ba \in \mathbb{R}^K$, classical bounds establish lower limits on the error in estimating $\ba^T\btheta$, given by $\mathbb{E}\left(\ba^T \hat{\btheta}(\by) - \ba^T \btheta\right)^2$, for a class of estimators $\hat{\btheta}(\by)$.  What we have left unspecified is the set of quantities we take the expectation over, and depending on this, the bounds fall into one of two categories:

\emph{Deterministic, but unknown, parameters:} One class of bounds do not use the prior distribution of $\btheta$, so that the parameter to be estimated $\btheta$ is best thought of as a deterministic but unknown quantity. The most popular such bound is the Cram\'{e}r Rao Bound (CRB). For the CRB, the expectation is taken over the conditional distribution $p(\by|\btheta)$, so that the bound is on $\mathbb{E}_{\by|\btheta} \left(\ba^T \hat{\btheta}(\by) - \ba^T \btheta\right)^2$.  The CRB typically depends on the parameter $\btheta$ and the most common version, which is what we use here, applies to estimators $\hat{\btheta}(\by)$ that are unbiased \footnote{An unbiased estimator $\hat{\btheta}(\by)$ is one which satisfies $\mathbb{E}_{\by|\btheta}\{\hat{\btheta}(\by)\} = \btheta$ for all $\btheta$}.

\emph{Bayesian bounds:} When we know the prior distribution $p(\btheta)$ from which $\btheta$ is chosen, we can incorporate this information into the bounds. Such bounds are called Bayesian bounds and, in these cases, the expectation is taken over the joint distribution $p(\by,\btheta) = p(\by|\btheta) p(\btheta)$. They establish lower limits on the Mean-Squared-Error (MSE) in estimating $\ba^T\btheta$, given by $\mathbb{E}_{\by,\btheta} \left(\ba^T \hat{\btheta}(\by) - \ba^T \btheta\right)^2$. Among the Bayesian bounds, we are primarily concerned with the Ziv-Zakai bound (ZZB) (we also briefly describe a version of the CRB, called the Bayesian CRB). Neither of these bounds (ZZB/BCRB) require the estimator to be unbiased.

The Ziv-Zakai Bound is known to be an accurate predictor of best possible estimation performance over a wide range of SNRs. Roughly speaking, it takes into account two sources of error: coarse error, when the estimate is not close to the true value of the parameter (essentially, making an error in hypothesis testing after binning the parameter space); and fine-grained error (the mean squared error from the true value when the estimate is in the right bin). At high SNR, the probability of the estimate falling into the wrong bin becomes negligible, and the Cram\'{e}r Rao bound (CRB), which characterizes only fine-grained error, provides an excellent prediction of performance, while being easier to compute than the ZZB.  We now state these bounds.

\subsection{Cram\'er Rao Bound\cite{van2013detection,book::2007::vantrees::bayesian}}
Let $\ba \in \mathbb{R}^K$. The variance of any unbiased estimator of $\ba^T\btheta$, given by $\mathbb{E}_{\by|\btheta} \left(\ba^T \hat{\btheta}(\by) - \ba^T \btheta\right)^2$, is lower bounded by $\ba^T F^{-1}(\btheta) \ba$, where $F(\btheta)$ is the Fisher Information Matrix (FIM). The $(m,n)$th element of the FIM is given by:
\begin{align}
F_{m,n}(\btheta) & = \mathbb{E}_{\by|\btheta}\left\{\frac{\partial\ln p(\by|\btheta)}{\partial\theta_m}\frac{\partial \ln p(\by|\btheta)}{\partial\theta_n}\right\}.
\end{align}
For parameter estimation in AWGN (\ref{eq:uncomp:meas:model:density}), this simplifies to \cite{van2013detection}
\begin{align}
F_{m,n}(\btheta) & = \frac{2}{\sigma^2}\Re\left\{\left(\frac{\partial \bs(\btheta)}{\partial\theta_m}\right)^H\frac{\partial \bs(\btheta)}{\partial\theta_n}\right\},\label{eq:FIM:AWGN}
\end{align}
where $\Re\{b\}$ denotes the real part of the complex number $b$. 

\subsection{Bayesian Bounds on Mean Square Error}
To describe the Bayesian bounds, it is convenient to define the MSE matrix, $R(\hat{\btheta})$ of the estimator $\hat{\btheta}(\by)$. The $m,n$-th element of the MSE matrix $R(\hat{\btheta})$ is given by $R_{m,n}(\hat{\btheta}) = \mathbb{E}_{\by,\btheta} \{(\hat{\theta}_m-\theta_m)(\hat{\theta}_n-\theta_n)\}$. For a vector $\ba\in\mathbb{R}^K$, the ZZB and BCRB provide bounds on $\mathbb{E}_{\by,\btheta} \big( \ba^T \hat{\btheta}(\by) - \ba^T \btheta\big)^2$ which is simply $\ba^T R(\hat{\btheta}) \ba$.

\subsubsection{Bayesian Cram\'er Rao Bound\cite{van2013detection,book::2007::vantrees::bayesian}}
For any weight vector $\ba \in \mathbb{R}^K$ and estimator $\hat{\btheta}(\by)$ (not necessarily unbiased),  the Bayesian Cram\'er Rao Bound (BCRB) lower bounds the MSE $\ba^T R(\hat{\btheta}) \ba$ by $\ba^T B^{-1} \ba$, where $B$ is the Bayesian Information Matrix (BIM).  The $(m,n)$th element of $B$ is given by:
\begin{equation}
B_{m,n} = \mathbb{E}_{\btheta} \left\{ F_{m,n}(\btheta) \right\} + \mathbb{E}_{\btheta}\left\{\frac{\partial\ln p(\btheta)}{\partial\theta_m}\frac{\partial \ln p(\btheta)}{\partial\theta_n}\right\}.
\label{eq:BIM:AWGN}
\end{equation}

\subsubsection{(Extended) Ziv-Zakai Bound\cite{Bell::EZZB::1997}}
Since the ZZB is not as widely used as the CRB, we provide a brief review in Appendix \ref{appendix:zzbReview}.  Here, we simply state the bound. The ZZB
bounds the MSE $\ba^T R(\hat{\btheta})\ba$ and, for the AWGN measurement model (\ref{eq:uncomp:meas:model}), it is given by:
\begin{align}
\ba^T R(\hat{\btheta})\ba  &\geq \frac{1}{2}\int_{0}^{\infty} \mathcal{V}\Bigg\{  \underset{\bdelta:\mathbf{a}^T \bdelta=h}{\max} \int_{\bphi\in\mathbb{R}^K}\!\!\!\!\! \left(p(\bphi) +\right.\nonumber\\
&\!\!\!\!\!\!\left. p(\bphi + \bdelta) \right)  f(\bphi,\bphi+\bdelta)\ d \bphi  \Bigg\} h\ dh \mc{\forall \hat{\btheta}(\by)}\label{eq:ZZB:bound}
\end{align}
where $\mathcal{V}\{\ \}$ is the valley filling operation, defined as $\mathcal{V}\{g(h)\}=\max_{r \geq 0}g(h+r)$, and $f(\btheta_{1},\btheta_{2})$ is the probability of error for the optimal detection rule in the following {\it hypothesis testing problem}:
\begin{align}
H_{1}\ :\ \by = \bs(\btheta_{1}) + \bz,\  \Pr(H_{1}) &= \frac{p(\btheta_{1})}{p(\btheta_{1}) + p(\btheta_{2})} \nonumber\\
H_{2}\ :\ \by =  \bs(\btheta_{2}) + \bz,\  \Pr(H_{2}) &= \frac{p(\btheta_{2})}{p(\btheta_{1}) + p(\btheta_{2})}.
\label{eq:ZZB:detection:problem}
\end{align}
Since $\bz \sim \mathcal{CN}(\mathbf{0},\sigma^{2}\mathbb{I}_M)$, this detection error probability is given by\cite{van2013detection}:
\small
\begin{align}
f(\btheta_1,\btheta_2) &=  \frac{p(\btheta_{1})}{p(\btheta_{1}) + p(\btheta_{2})}  Q\left( \frac{d(\btheta_1,\btheta_2)}{\sqrt{2}\sigma} + \frac{\sigma}{\sqrt{2}d(\btheta_1,\btheta_2)} \ln\frac{p(\bm{\theta}_1)}{p(\bm{\theta}_2)}\right)\nonumber\\
 &\!\!\!\!\!\!\!\!\!\!  +  \frac{p(\btheta_{2})}{p(\btheta_{1}) + p(\btheta_{2})}  Q\left(\frac{d(\btheta_1,\btheta_2)}{\sqrt{2}\sigma}	-\frac{\sigma}{\sqrt{2}d(\btheta_1,\btheta_2)}\ln\frac{p(\bm{\theta}_1)}{p(\bm{\theta}_2)}\right).\label{eq:ZZB:detection:probability}
\end{align}
\normalsize
In the above expression, $Q(\ )$ stands for the CCDF of the standard normal distribution $\mathcal{N}(0,1)$ and 
\begin{equation}
d(\btheta_1,\btheta_2) = \Vert\bs(\btheta_1)-\bs(\btheta_2)\Vert.
\end{equation}

\noindent \textbf{Remark:}  While the expression for the ZZB is complicated,  we only need two simple observations to prove the result we are interested in: 

$\bullet$ With compressive measurements, the signal manifold $\bs(\btheta) = \bA \bx(\theta)$ and the measurement matrix $\bA$ enters the ZZB {\it only} through the pairwise SNRs $d^2(\btheta_1,\btheta_2)/\sigma^2$.

$\bullet$ The minimum probability of detection error $f(\btheta_1,\btheta_2)$ for the binary hypothesis testing problem (\ref{eq:ZZB:detection:problem}) is a non-increasing function of the pairwise SNR $d^2(\btheta_1,\btheta_2)/\sigma^2$. 

We revisit these observations in Section \ref{sec:isometryBoundRelation}. 

\subsection{Threshold behavior of ZZB}
\label{sub-sec:ZZB:Thresh}
The ZZB typically exhibits a {\it threshold behavior} with SNR \cite{book::2007::vantrees::bayesian}. When the SNR is very low, the measurements carry little information about the parameters we wish to estimate.  Since the ZZB accounts for errors of ``all magnitudes", it is usually large (depending primarily on the prior $p(\btheta)$) and insensitive to small changes in SNR in this regime. However,  at high SNRs, the variation of the ZZB with SNR is predictable. When the SNR and the ZZB are both expressed on a logarithmic scale, the ZZB falls off linearly with SNR,  provided that the SNR is above a certain value, which is called the (asymptotic) {\it ZZB threshold} \cite{Bell::EZZB::1997}. When the SNR exceeds the ZZB threshold, ``large'' estimation errors are unlikely, which is exactly when we would declare estimation of a continuous-valued parameter to be successful. 

\section{Relating the isometries to estimation bounds}
\label{sec:isometryBoundRelation}

We are now ready to relate the estimation error bounds for the compressive estimation problem to the corresponding bounds when we make all $N$ measurements, provided that the compressive measurement matrix $\bA$ satisfies appropriate isometry conditions.

Consider the general problem of estimating $\btheta$ from $L$ measurements
\begin{equation}
\by = \bB \bx(\btheta ) + \mathbf{z}, \quad \btheta \in \Theta
\label{eq:generalModel}
\end{equation}
where $\bB$ is any $L \times N$ complex-valued matrix and $\mathbf{z} \sim \mathcal{CN}(\mathbf{0},\sigma^{2}\mathbb{I}_L)$. The compressive estimation problem is subsumed in this model (obtained by setting $\bB = \bA$, whose elements are chosen i.i.d. from a zero-mean distribution of variance $1/N$ for which concentration results of the form (\ref{eq:concentration:projection}) are available), as is the conventional problem of estimating $\btheta$ from all $N$ measurements (obtained by setting $\bB = \mathbb{I}_{N}$, the $N\times N$ identity matrix). Note that, in both these cases, the per-measurement SNR $(1/L)\sum)_{k=1}^{k=L}\mathbb{E}|y_{k}|^{2}/\sigma^{2}$ is the same, since the rows of $\bA$ have unit norm in expectation.

We prove two theorems that connect the fundamental estimation-theoretic bounds to the isometries defined in the previous section. 
First, we make a connection between the ZZB and the pairwise isometry property. As we observed in the remark under the statement of the ZZB, for the manifold $\bs(\btheta) = \bB\bx(\btheta)$, the ZZB depends on the matrix $\bB$ only through the set of pairwise SNRs $\Vert \bB \bx(\btheta_{1}) - \bB \bx(\btheta_{2}) \Vert^{2}/\sigma^{2}~\forall \btheta_{1},\btheta_{2} \in \Theta$. When the compressive measurement matrix $\bA$ satisfies the pairwise isometry property (\ref{eq:PIP}), the pairwise SNRs with $\bB = \bA$ are approximately $M/N$ times the corresponding values with $\bB = \mathbb{I}_N$. Thus, the ZZB with compressive measurements is approximately the same as the ZZB with all $N$ measurements, but at an SNR penalty of $M/N$. Theorem \ref{thm:ZZB:relation} proves this intuition rigorously. 

Likewise, we can connect the CRB to the tangent-plane isometry property. We can show that the CRB depends on the measurement matrix only through norms of the vectors $\bB\sum_m a_m (\partial \bx(\btheta )/\partial \theta_m)$. Thus, if $\bA$ satisfies the tangent-plane isometry (\ref{eq:tangent_plane:isometry}), the CRB with $M$ compressive measurements is approximately equal to the CRB with all $N$ measurements, but at an SNR that is lower by $M/N$. We prove this in Theorem \ref{thm:CRB:relation}.

While the connections established here between estimation-theoretic bounds and the corresponding isometries apply generally to compressive estimation in AWGN, showing that these isometries indeed hold requires a problem-specific analysis, as we illustate for sinusoidal mixtures in later sections. As with standard compressed sensing, the goal of such analyses is to characterize the number of measurements required for such isometries to hold with high probability for random measurement matrices.

\subsection{Cram\'{e}r Rao Bound}
\label{sub-sec:CRB:relation}

Let $F(\bB,\btheta)$ denote the Fisher Information Matrix for the measurement model (\ref{eq:generalModel}). For this measurement model the expression for FIM is given by (\ref{eq:FIM:AWGN}) with $\bs(\btheta) = \bB\bx(\btheta)$:
\begin{dmath}
F_{m,n}(\bB,\btheta)  = \frac{2}{\sigma^2}\Re\left\{\left(\bB\frac{\partial \bx(\btheta)}{\partial\theta_m}\right)^H \bB\frac{\partial \bx(\btheta)}{\partial\theta_n}\right\}.\label{eq:FIM:AWGN:iso} 
\end{dmath}

\begin{theorem}\label{thm:CRB:relation}
Let $\bA$ be an $M\times N$ measurement matrix which satisfies the tangent plane $\epsilon$-isometry property (\ref{eq:tangent_plane:isometry}) for the signal manifold $\bx(\btheta)$. Then, the Fisher Information Matrix $F(\bA,\btheta)$, with compressive measurements (\ref{eq:compressive:measurement:model}) is related to the FIM with all $N$ measurements as follows:
\begin{equation}
\begin{array}{c}
F(\bA,\btheta)\preceq F\left(\sqrt{\frac{M}{N}}(1+\epsilon)\mathbb{I}_N,\btheta\right)\\
F(\bA,\btheta)\succeq F\left(\sqrt{\frac{M}{N}}(1-\epsilon)\mathbb{I}_N,\btheta\right)
\end{array}\ \forall\btheta\in\Theta .\label{eq:FIM:relations:SNRPenalty}
\end{equation}
\end{theorem}
\begin{IEEEproof}
Consider the quadratic form $\ba ^TF(\bB,\btheta)\ba $ for any $\ba =[a_1\ \cdots\ a_K]^T\in\mathbb{R}^K$. We see that
\begin{dmath}
	\ba ^TF(\bB,\btheta)\ba  =  \frac{2}{\sigma^2}\left\Vert\bB\sum_m a_m\frac{\partial \bx(\btheta )}{\partial \theta_m}\right\Vert^2. \label{eq:FIM:quad-form}
\end{dmath}
Since the compressive measurement matrix $\bA$ satisfies the tangent plane $\epsilon$-isometry property (\ref{eq:tangent_plane:isometry}) for the signal model $\bx(\btheta)$, we have that for all $\btheta\in\Theta$ and $\ba\in\mathbb{R}^K$,
\begin{equation}
\left\Vert\bA\sum_m a_m\frac{\partial \bx(\btheta )}{\partial \theta_m}\right\Vert^2 \leq \frac{M}{N} (1+\epsilon)^{2}\left\Vert \sum_m a_m\frac{\partial \bx(\btheta )}{\partial \theta_m}\right\Vert^2.
\end{equation}
Multiplying both sides by $2/\sigma^{2}$, we see that the LHS is $\ba ^TF(\bA,\btheta)\ba $, while the RHS corresponds to $\ba ^T F\left(\sqrt{{M}/{N}}(1+\epsilon)\mathbb{I}_N,\btheta\right) \ba $. Therefore, we have that $\forall\btheta\in\Theta$,
\begin{align}
\ba ^TF(\bA,\btheta)\ba  &\leq \ba ^{T} F\left(\sqrt{{M}/{N}}(1+\epsilon)\mathbb{I}_N,\btheta\right) \ba,\ \forall \ba \in\mathbb{R}^K.\label{eq:CRB:Proof:UB}
\end{align}
This establishes the required upper bound on $F(\bA,\btheta)$. The proof for the lower bound is analogous. 
\end{IEEEproof}
\subsection{Bayesian Cram\'{e}r Rao Bound}
Let $B(\bB)$ denote the Bayesian Information Matrix for the measurement model (\ref{eq:generalModel}). Let $p(\btheta)$ be the prior on $\btheta$. For this measurement model the expression for BIM is given by (\ref{eq:BIM:AWGN}) with $\bs(\btheta) = \bB\bx(\btheta)$:
\begin{equation}
B_{m,n}(\bB)  = \mathbb{E}_{\btheta} \left\{ F_{m,n}(\bB,\btheta) \right\} + \mathbb{E}_{\btheta}\left\{\frac{\partial\ln p(\btheta)}{\partial\theta_m}\frac{\partial \ln p(\btheta)}{\partial\theta_n}\right\}.\label{eq:BIM:AWGN:iso} 
\end{equation}

\begin{corollary}[of Theorem \ref{thm:CRB:relation}]
Let $\bA$ be an $M\times N$ measurement matrix which satisfies the tangent plane $\epsilon$-isometry property (\ref{eq:tangent_plane:isometry}) for the signal manifold $\bx(\btheta)$. Then, the Bayesian Information Matrix $B(\bA)$ with compressive measurements (\ref{eq:compressive:measurement:model}) is related to the BIM with all $N$ measurements as follows:
\begin{equation}
 B\left(\sqrt{\frac{M}{N}}(1-\epsilon)\mathbb{I}_N\right) \preceq B(\bA) \preceq B\left(\sqrt{\frac{M}{N}}(1+\epsilon)\mathbb{I}_N\right) \label{eq:BIM:relations:SNRPenalty}
\end{equation}
\end{corollary}
\begin{IEEEproof}
Let $\ba\in\mathbb{R}^K$. We see that $\ba^T B(\bA)\ba$ depends on the measurement matrix $\bA$ only through quadratic forms of the FIM i.e., $\ba^T F(\bA,\btheta) \ba$. When the tangent plane isometry condition (\ref{eq:tangent_plane:isometry}) is satisfied, we have from Theorem \ref{thm:CRB:relation} that $ \ba^T F(\bA,\btheta) \ba$ is bounded by $\ba^T F(\sqrt{M/N}(1\pm\epsilon)\mathbb{I}_N,\btheta) \ba$ for all $\ba, \btheta$. It immediately follows that the quadratic forms of $B(\bA)$ are bounded by the corresponding quadratic forms of $B(\sqrt{M/N}(1\pm\epsilon)\mathbb{I}_N)$.
\end{IEEEproof}

\subsection{Ziv-Zakai Bound}
\label{sub-sec:ZZB}
Let $Z(\bB,\ba)$ denote the ZZB corresponding to the Mean-Squared-Error in estimating $\ba^T\btheta$ for the measurement model (\ref{eq:generalModel}). The expression for $Z(\bB,\ba)$ is given by the right hand side of (\ref{eq:ZZB:bound}), with $d(\btheta_1,\btheta_2) = \Vert \bB \bx(\btheta_1) - \bB \bx(\btheta_2) \Vert$ (obtained by setting $\bs(\btheta) = \bB \bx(\btheta)$). 

Note that in (\ref{eq:ZZB:bound}), $f(\btheta_1,\btheta_2)$ is the probability of detection error for the hypothesis testing problem (\ref{eq:ZZB:detection:problem}) with $\bs(\btheta) = \bB\bx(\btheta)$. We capture the dependence of this probability on the matrix $\bB$ by defining $g(\bB,\btheta_1,\btheta_2) = f(\btheta_1,\btheta_2)$ when $\bs(\btheta) = \bB \bx(\btheta)$.

\begin{theorem}\label{thm:ZZB:relation}
Let $\bA$ be an $M\times N$ measurement matrix which satisfies the pairwise $\epsilon$-isometry property (\ref{eq:PIP}) for the signal manifold $\bx(\btheta )$. Then, the ZZB $Z(\bA,\ba)$, with the compressive measurements in (\ref{eq:compressive:measurement:model}), is related to the ZZB with all $N$ measurements as
\small
\begin{equation}\label{eq:ZZB:relations:SNRPenalty}
 Z\left(\sqrt{\frac{M}{N}}(1+\epsilon)\mathbb{I}_N ,\ba \right) \leq Z\left(\bA,\ba \right) \leq Z\left(\sqrt{\frac{M}{N}}(1-\epsilon)\mathbb{I}_N ,\ba \right).
\end{equation}
\normalsize
\end{theorem}
\begin{IEEEproof}
As we observed in the remark at the end of the definition of the ZZB,  $g(\bB,\btheta_1,\btheta_2)$ is a non-increasing function of the pairwise SNR $\Vert\bB\bx(\btheta_1)-\bB\bx(\btheta_2)\Vert^2/\sigma^2$. When $\bA$ satisfies the pairwise $\epsilon$-isometry property (\ref{eq:PIP}), we can bound all the pairwise SNRs as follows:
\begin{align}
{\Vert\bA\bx(\btheta_1)-\bA\bx(\btheta_2)\Vert^2}/{\sigma^2} &\leq {\frac{M}{N}} (1+\epsilon)^2 {\Vert \bx(\btheta_{1}) - \bx(\btheta_{2})\Vert^2}/{\sigma^2}\nonumber\\&\quad\quad\quad\quad\forall \btheta_1,\btheta_2\in\Theta.
\end{align}
Combining these facts, we get $g(\bA,\btheta_{1},\btheta_{2}) \geq g(\sqrt{M/N}(1+\epsilon)\mathbb{I}_{N},\btheta_{1},\btheta_{2})$, which is the probability of detection error with all $N$ measurements, but at an SNR penalty of $(M/N) (1+\epsilon)^2$. Substituting these pointwise bounds in the expression for $Z(\bA,\ba)$, we have that $ Z\left(\bA,\ba \right) \geq Z\left(\sqrt{\frac{M}{N}}(1+\epsilon)\mathbb{I}_N ,\ba \right)$. The other inequality can be proved similarly.
\end{IEEEproof}

\subsection{Number of measurements needed}
\label{sec:GeneralBounds,sub-sec:NumMeas}
These theorems show that, when the compressive measurement matrix $\bA$ satisfies the pairwise isometry property, the CRB and the ZZB are well approximated by $\ba^T F^{-1}(\sqrt{M/N} \mathbb{I}_N,\btheta) \ba$ and $Z(\sqrt{M/N} \mathbb{I}_N,\ba)$ respectively (for any $\ba$). Thus, the estimation performance with the measurement matrix $\bB = \bA$ is roughly the same as that with $\bB = \sqrt{M/N} \mathbb{I}_N$ (all $N$ measurements, but with the signal component scaled by $\sqrt{M/N}$). Note that observations with $\bB = \sqrt{M/N} \mathbb{I}_N$ and per-sample noise variance $\sigma^2$ are {\it equivalent} to observations $\bB = \mathbb{I}_N$ (conventional measurements) but with an increased per-sample noise variance $\sigma^2 (N/M)$ (easily seen by multiplying the observations with $\bB = \sqrt{M/N} \mathbb{I}_N$ by $\sqrt{N/M}$). Putting these observations together, we get a simple procedure for estimating the number of measurements $M$ required for successful compressive estimation:

(1) For the case when we make all $N$ measurements, $\by = \bx(\btheta) + \bz$ with $\bz \sim \mathcal{CN}(0,\sigma^2 \mathbb{I}_N)$, compute the ZZB as a function of $\sigma^2$. Find the ZZB threshold as described in Section \ref{sec:ParameterEstimationReview} (the value of $\sigma^2$ \textit{below} which $\log \text{ZZB}$ falls off linearly with $\log \sigma^2$). Denote this threshold by $\sigma_{\text{t}}^2$.

(2) Making $M$ compressive measurements $\by = \bA \bx(\btheta) + \bz$ with $\bz \sim \mathcal{CN}(0,\sigma_0^2 \mathbb{I}_M)$ is roughly equivalent to making the observations $\tilde{\by} = \bx(\theta) + \tilde{\bz}$ with $\tilde{\bz} \sim \mathcal{CN}(0,\sigma_0^2 (N/M) \mathbb{I}_M)$ when $\bA$ satisfies the pairwise isometry property. Thus, the number of measurements needed for successful compressive estimation is given by:
\begin{equation}
\sigma_0^2 \frac{N}{M} < \sigma_{\text{t}}^2 \quad \text{or} \quad M > N \left(\frac{\sigma_0^2 }{\sigma_{\text{t}}^2 }\right)
\end{equation}
We reiterate that the above SNR criterion is not the only condition for successful compressive estimation: the number of measurements $M$ must be large enough for the matrix $\bA$ to satisfy the pairwise isometry property, so that we can invoke the SNR penalty arguments.

In the next section, we illustrate these ideas by considering the example of frequency estimation of a single sinusoid. But before that, we comment on the generality of the model we have considered so far.

\noindent \textbf{Remarks on model generality:} While we describe our results in the context of the measurement model (\ref{eq:compressive:measurement:model}), they extend easily to variants commonly encountered in the compressed sensing literature, two of which we now discuss.

$\bullet$ For applications such as Direction of Arrival (DoA) estimation using large arrays \cite{allerton:tracking}, compressive measurements are acquired sequentially in time and every measurement is corrupted by independent {\it measurement} noise. Thus, the measurements satisfy
\begin{equation}
y_l  = \bw_l^T\left(\bx(\btheta) + \tilde{\bz}_l\right) = \bw_l^T\bx(\btheta) +z_l,\label{eq:meas:one:alt}
\end{equation}
where $\tilde{\bz}_l\sim\mathcal{CN}\left(\mathbf{0},\sigma^2\mathbb{I}_N\right)$ and $z_l = \bw_l^T\tilde{\bz}_l\sim\mathcal{CN}(0,\sigma^2\Vert\bw_l\Vert^2)$. The key point here is that $\tilde{\bz}_1,\dots,\tilde{\bz}_M$ are i.i.d. and as a result $z_1,\dots,z_M$ are independent. Letting $\bA$ denote the matrix with rows $\bw_l^T$, $\by = [y_1\ \cdots\ y_M]^T$ and $\bz = [z_1\ \cdots\ z_M]^T$, we have:
\begin{align}
\by &= \bA\bx(\btheta) + \bz \mc{\bz\sim\mathcal{CN}(\mathbf{0},\sigma^2\bK_1)},
\end{align}
where $\bK_1$ is a diagonal matrix whose diagonal entries are $\Vert\bw_l\Vert^2,~l=1,\dots,M$.

$\bullet$ For other applications, when we have access to a {\it single} noisy version of $\bx(\btheta)$ and compressive measurements are merely used as a {\it dimensionality reduction} tool, we have
\begin{align}
\by	& = \bA\left(\bx(\btheta) + \tilde{\bz}\right) \mc{\tilde{\bz}\sim\mathcal{CN}(\mathbf{0},\sigma^2\mathbb{I}_N)}.\label{eq:meas:two}
\end{align}
The same equation holds for the case when there are errors in modeling the manifold $\bx(\btheta)$ (given by $\tilde{\bz}$) and we make $M$ sequential {\it noiseless} projections.
Letting $\bz = \bA\tilde{\bz}$ we have
\begin{align}
\by	& =\bA\bx(\btheta) +\bz \mc{\bz\sim\mathcal{CN}(\mathbf{0},\sigma^2\bK_2)},
\end{align}
where $\bK_2 = \bA \bA^H$. 

Neither $\bK_1$ and $\bK_2$ are the identity matrix, hence these measurement models do not fit directly into the framework in (\ref{eq:compressive:measurement:model}).  However, we can extend our results easily to these models by considering the whitened observations $\tilde{\by}_i = \bK_i^{-1/2} \by,~i = 1,2$, and establishing bounds on the singular values of $\bK_i$. When the elements of $\bA$ are chosen from a zero-mean distribution of variance $1/N$ (for which concentration results of the form (\ref{eq:concentration:projection}) are available), the singular values of $\bK_i$ concentrate around $1$. As a result, an $\epsilon$-isometry (tangent plane or pairwise) for $\bA$ can be shown to translate to a mildly weaker $\epsilon_{\text{eff},i}$-isometry ($\epsilon_{\text{eff},i} \geq \epsilon$) for $\bA_{\text{eff},i}=\bK_i^{-1/2}\bA$, the effective measurement matrix for the whitened measurements $\tilde{\by}_i$.  All of our results now apply by simply replacing $\epsilon$ with $\epsilon_{\text{eff},i}$. This equivalence of the measurement model (\ref{eq:meas:two}) and the general compressive model (\ref{eq:compressive:measurement:model}) has also been investigated in detail in \cite{Eldar_Noise_folding_2011}. The proof for the conditioning of both $\bK_1$ and $\bK_2$ involves using the concentration result (\ref{eq:concentration:projection}) for $\sqrt{N/M}\bA^H$ (see \cite{vershynin_introduction_2010} for $\bK_2$). 

The concentration results for the singular values of $\bK_2 = \bA\bA^H$ (which are the square of the singular values of $\bA^H$) needs $N$ to be somewhat larger than $M$. This is not an issue, since this is the regime of interest for compressive estimation. The diagonal matrix $\bK_1$, on the other hand, is well-conditioned for much larger values of $M$ (potentially larger than $N$). 

\section{Designing compressive estimation strategies} 
\label{sec:frequencyestimation:bounds}

In this section, we illustrate, using the example of frequency and phase estimation for a single sinusoid, how to apply the preceding results to design compressive estimation strategies.  We describe an algorithm which attains the CRB given ``enough'' compressive measurements, and show how to determine how many measurements are enough, based on the threshold behavior of the ZZB. We implicitly assume that we have enough measurements for the appropriate isometries to hold; detailed analytical characterization of the number of measurements required for this purpose is deferred to later sections.

The measurements are given by
\begin{equation}
\by = e^{j\phi} \bB \bx(\omega) + \bz
\label{eq:sineModel}
\end{equation}
where $\bx(\omega) = \left[e^{-j\omega(N-1)/2}\ e^{-j\omega(N-3)/2}\ \cdots\ e^{j\omega(N-1)/2}\right]^T$ is an $N$-dimensional sinusoid with frequency $\omega$, $\phi$ is its phase, $\bB$ is an $L\times N$ complex valued measurement matrix and $\bz \sim \mathcal{CN}(0,\sigma^{2}\mathbb{I}_L)$. The parameters to be estimated $\phi$ and $\omega$ are both distributed uniformly over $[0,2 \pi]$. Note that there is a slight change in notation from the previous section. Earlier, we denoted the parameter to be estimated by $\btheta = \left[\omega\ \phi \right]^{T}$ and the signal manifold $\bx(\btheta) = e^{j \phi}  \left[e^{-j\omega(N-1)/2}\ e^{-j\omega(N-3)/2}\ \cdots\ e^{j\omega(N-1)/2}\right]^T$. We now separate the contributions from the phase and frequency and use $\bx(\omega)$ to denote a sinusoid with frequency $\omega$ and zero phase ($\phi = 0$).

When we make all $N$ measurements (setting $\bB = \mathbb{I}_{N}$ in (\ref{eq:sineModel})), the CRB is well known \cite{rife:boorstyn:CRB:freq}. The FIM in estimating $\btheta = [\omega ~\phi]$ is 
\begin{equation}
F(\mathbb{I}_{N},\btheta)=\frac{2}{\sigma^2}\left[
\begin{matrix}
N(N^{2}-1)/12& 0\\ 
0&N
\end{matrix} 
\right]\ \forall\btheta.
\label{eq:FIMfull}
\end{equation}
In particular, the CRB on the variance of the frequency estimate (computed as $\ba^{T} F^{-1}(\mathbb{I}_{N},\btheta) \ba$ with $\ba = \left[1\ 0\right]$) is $\text{CRB}(\mathbb{I}_{N},\btheta) = 6\sigma^2/(N(N^2-1))$. Note that the CRB is independent of $\btheta$. 

We must be careful in computing the ZZB because the noiseless signal is a periodic function (with period $2 \pi$) of both the phase and the frequency.
Thus, the errors in estimating these parameters must be appropriately defined (i.e., the difference between $0$ and $2\pi-\epsilon$ is $\epsilon$ for small $\epsilon$). The ZZB on the ``periodic-MSE'' of the frequency estimate is given by (using (27) in \cite{Basu::ZZB::Periodic::2000})
\begin{align}
Z(\mathbb{I}_{N},\ba) = &  \int_{0}^{\pi}\!\!\!  \underset{\phi^{\prime}\in[0,2\pi]}{\max}\  Q\left(\frac{\Vert\mathbf{x}(0)-e^{j\phi^{\prime}}\mathbf{x}(h)\Vert}{\sqrt{2}\sigma}\right)h\ dh\nonumber\\
		        =& \int_{0}^{\pi} Q\left(\sqrt{\frac{N}{\sigma^2}\left(1-\left|\frac{\sin(Nh/2)}{N\sin(h/2)}\right|\right)}\right)h\ dh.
\end{align}
Suppose now that we make $M$ compressive measurements (setting $\bB = \bA$), choosing $M$ large enough so that the measurement matrix $\bA$ satisfies the pairwise $\epsilon$-isometry property for the $\left\{e^{j\phi}\bx(\omega)\right\}$ signal model (the number of compressive measurements needed to establish pairwise isometries for this signal model is analytically characterized in Section \ref{sub:sec:SingleSinusoid}). Then, from Section \ref{sec:isometryBoundRelation}, we know that the Fisher information with compressive measurements $F\left(\bA,\btheta \right)$ is well-approximated by $F\left(\sqrt{M/N}~\mathbb{I}_{N},\btheta\right)$, the Fisher information with all $N$ measurements at an $M/N$ SNR penalty. Given that we know $F(\mathbb{I}_{N},\btheta)$, computing $F(\sqrt{M/N}~\mathbb{I}_{N},\btheta)$ is easy: we simply replace $\sigma^{2}$ in (\ref{eq:FIMfull}) by $\sigma^{2}(N/M)$.

When $\bA$ satisfies the pairwise isometry property, we can show that the ZZB with periodic-MSE also satisfies Theorem \ref{thm:ZZB:relation}. Therefore, the above arguments regarding the increase in the noise level by a factor of $N/M$ hold true for the ZZB with periodic distortion too.  Thus, we get the CRB and the ZZB with compressive measurements to be 
\begin{equation}
\text{CRB}(\bA,\btheta) \approx \text{CRB}(\sqrt{M/N}~\mathbb{I}_{N},\btheta) =  6 \sigma^{2}/(M(N^{2}-1))\ \forall\btheta
\end{equation}
\begin{eqnarray}
\!\!\!\!\!\!\!\!Z(\bA,\ba) &\!\!\!\!\approx&\!\!\!\! Z(\sqrt{M/N}~\mathbb{I}_{N},\ba)\nonumber\\
&\!\!\!\! = &\!\!\!\! \int_{0}^{\pi}  \!Q\left(\sqrt{\frac{M}{\sigma^2}\left(1-\left|\frac{\sin(Nh/2)}{N\sin(h/2)}\right|\right)}\right) h\ dh.
\end{eqnarray}

\begin{figure*}
\centering
\begin{minipage}{0.97\columnwidth}
\centering
\includegraphics[trim = 0.375in 0in 0.375in 0.375in, clip=true, width=0.83\columnwidth,height=0.63\columnwidth]{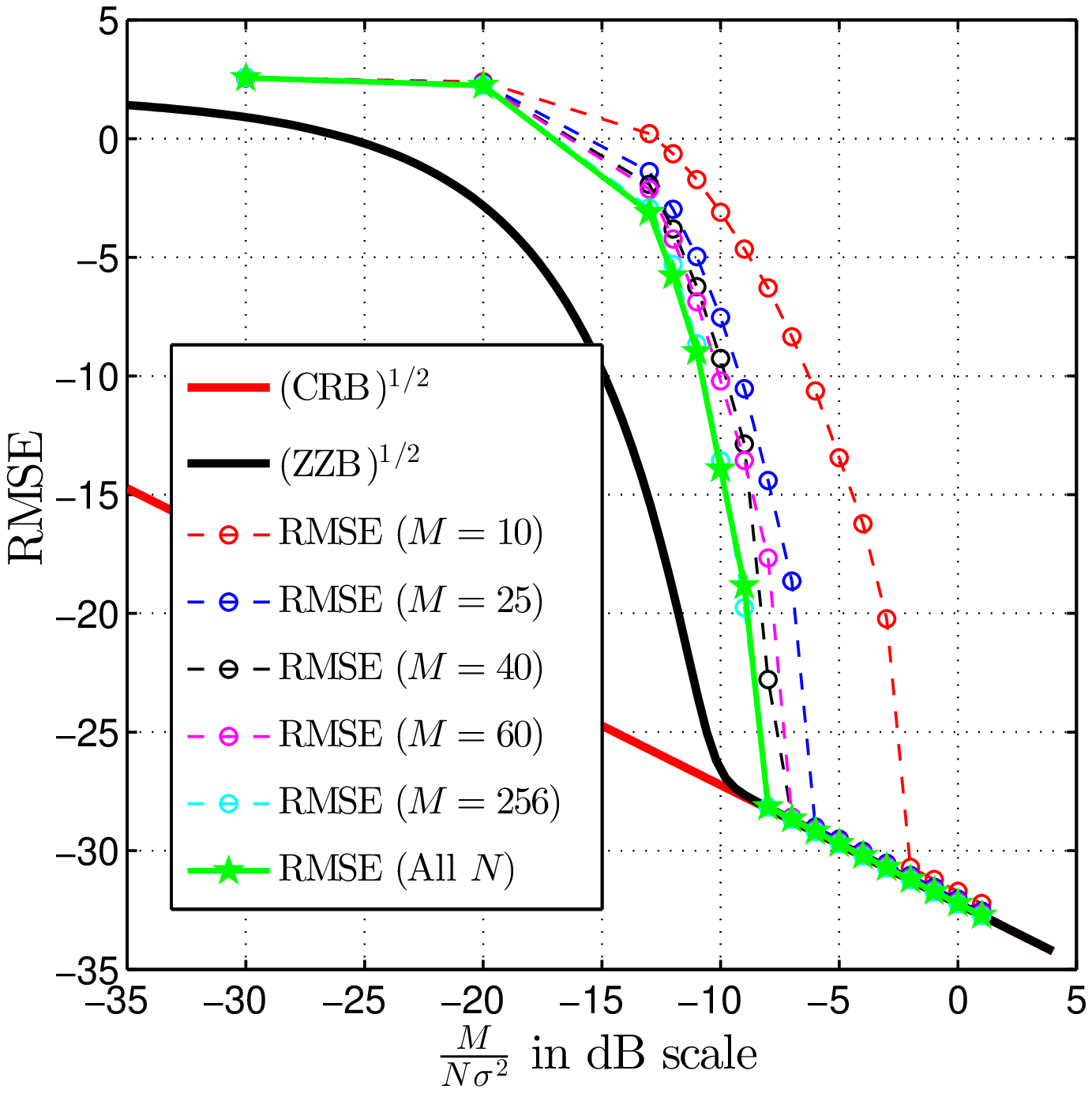}
\caption{RMSE in dB scale for 5 compressive measurement matrices ($\bB = \bA$) with $M= 10, 25, 40, 60, 256$ and the all $N$ measurements case ($\bB = \mathbb{I}_N$) plotted against effective per sample SNR $M/(N\sigma^2)$. Overlaid are plots of $\sqrt{\text{CRB}}$ and $\sqrt{\text{ZZB}}$ for all $N$ measurements ($\bB = \mathbb{I}_N$) corresponding to this effective SNR. The length of the sinusoid $\bx(\omega)$ is $N = 256$.}
\label{fig:RMSE}
\end{minipage}
\qquad
\begin{minipage}{0.97\columnwidth}
\centering
\includegraphics[trim = 0.275in 0in 0in 0.1in, clip=true, width=0.71\columnwidth,height=0.63\columnwidth]{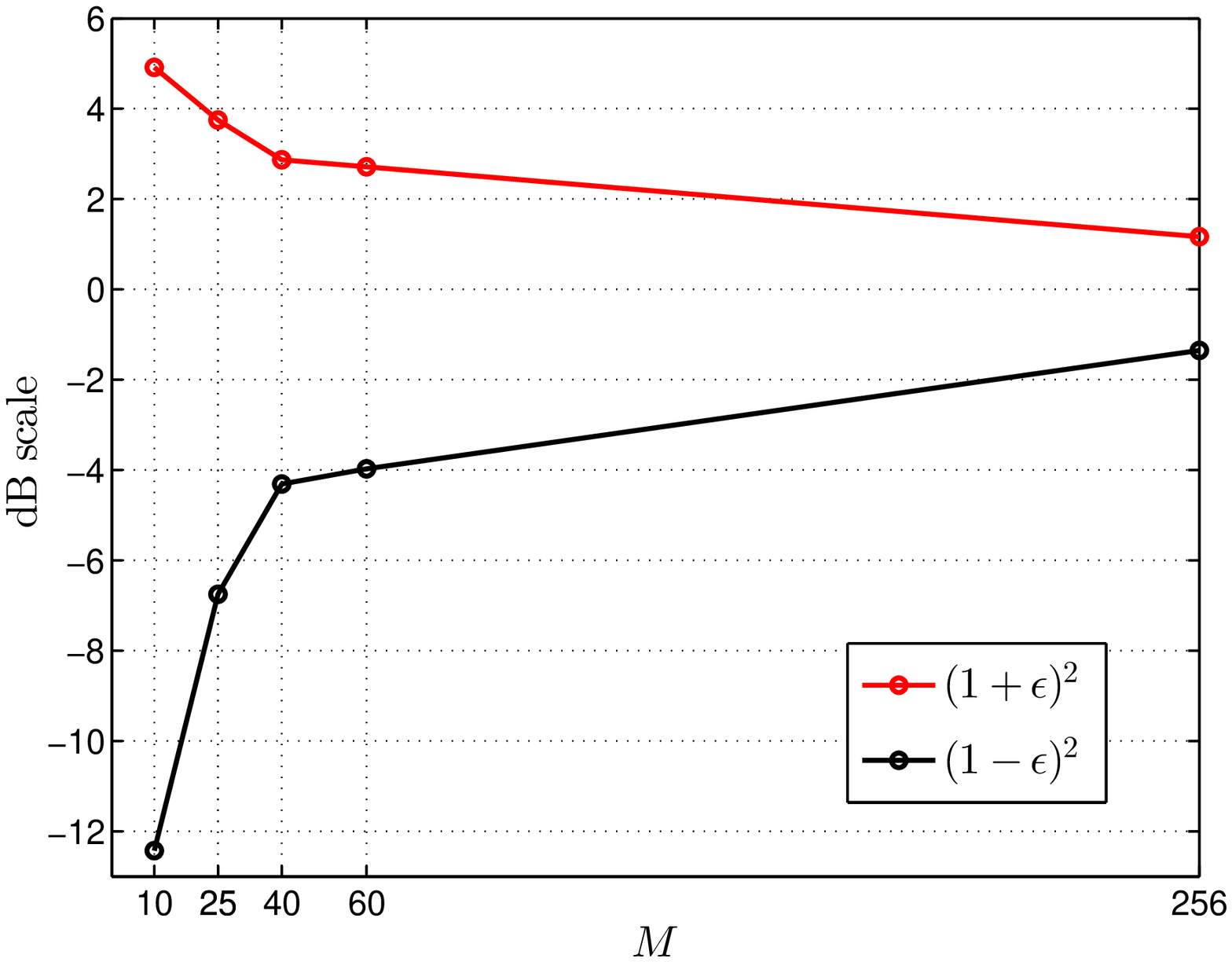}
\caption{\textit{Bounds} on pairwise SNR variation due to pairwise isometry constant $\epsilon$ (\ref{eq:PIP}) for the compressive measurement matrices used in Fig.~\ref{fig:RMSE}. Isometry constant $\epsilon$ corresponds to the manifold $\{g e^{j\phi}\bx(\omega)\}$ where $g\in\mathbb{R}^+$ and $\phi,\omega\in[0,2\pi]$.}
\label{fig:iso_Consts}
\end{minipage}

\end{figure*}

We now illustrate how to predict the number of measurements needed for successful compressive estimation based on the threshold behavior of the ZZB.
Consider frequency estimation of a $N=256$ sinusoid from all $N$ measurements ($\bB=\mathbb{I}_N$) at a noise level $\sigma^{2}$. In Fig.~\ref{fig:RMSE}, we plot the CRB and the ZZB for this estimation problem as a function of the per-measurement SNR $\overset{\triangle}{=}1/\sigma^2$. For SNRs that are smaller than $-30$dB, we see from Fig.~\ref{fig:RMSE} that the ZZB is insensitive to changes in SNR, unlike the CRB which exhibits a linear falloff for all SNRs. However, when the SNR exceeds $-10$dB, the ZZB exhibits a linear falloff with SNR. 

If we now make $M$ compressive measurements ($\bB = \bA$), the results of the previous section tell us that the effective SNR is given by
$(1/\sigma^2)(M/N)$.  We expect ``good'' estimation performance when this effective SNR exceeds the ZZB threshold, which translates
to the following rule of thumb for the number of compressive measurements required: 
\begin{equation} \label{eq:number_measurements_zzb}
M > N\sigma^2  \times \textrm{ZZB threshold SNR},
\end{equation}
Note that the ZZB threshold is computed for the {\it original} system with all $N$ measurements ($\bB = \mathbb{I}_N$), independent of the compressive measurement matrix $\bA$ and the noise level $\sigma^2$. 
For our specific example of a sinusoid of length $N=256$, the preceding prescription translates to $M > N \sigma^2/10$, since the ZZB threshold is $-10$ dB.

We now describe an algorithm whose performance closely follows these predictions: the algorithm approaches the CRB (for a given effective SNR) when
the effective SNR exceeds the \textit{ZZB threshold}. This illustrates the efficiency of the algorithm, as well as the accuracy of our design guideline of ``sufficient effective SNR." 

\noindent \textbf{Algorithm:} Suppose that for the purposes of algorithm design, we ignore the fact that the unknown phase rotation $e^{j \phi}$ has unit amplitude and estimate the complex gain $g$ and the frequency $\omega$ according to the model
\begin{equation}
\by = g \bB \bx(\omega) + \bz,\quad\bz\sim\mathcal{CN}(\mathbf{0},\sigma^2\mathbb{I}).
\end{equation}
The ML estimates of the gain and frequency $(\hat{g},\hat{\omega})$ are obtained by optimizing the function
\begin{equation}
S(g,\omega) = \Re\left\{\mathbf{y}^H g\bB\mathbf{x}(\omega)\right\}-0.5{|g|^2\left\Vert\bB\mathbf{x}(\omega)\right\Vert^2},
\end{equation}
over $g \in \mathbb{C}, \omega \in \left[0,2 \pi \right]$ and $\Re\{a\}$ denotes the real part of the complex number $a$. Performing a direct optimization over $g$ and $\omega$ is difficult. Therefore, we resort to a two stage procedure, consisting of a detection phase and a refinement phase, which we describe now. 

(i) \textit{Detection phase}: First, we notice that for any $\omega$, the optimizing $g$ is given by $\left(\bB \bx(\omega)\right)^{H} \by/\Vert \bB \bx(\omega) \Vert^{2}$. Substituting this in the cost function $S(g,\omega)$, we see that the ML estimate of the frequency $\hat{\omega}$ should optimize $G(\omega)=\max_{g\in\mathbb{C}} S(g,\omega)=0.5|\mathbf{y}^H \bB\mathbf{x}(\omega)|^2/\Vert\bB\mathbf{x}(\omega)\Vert^2$. We obtain a coarse frequency estimate by discretizing the frequencies uniformly into a set $F = \{0, 2\pi/(4N), \ldots, 2 \pi (4N-1)/(4N)\}$ of size $4N$ and then choosing $q^\star\in F$ that maximizes $G(q), q\in F$. Since the frequency estimation error is substantial (on the order of $1/N$), we call this the detection phase. The gain estimate is given by $\hat{g} = \left(\bB \bx(q^{\star})\right)^{H} \by/\Vert \bB \bx(q^{\star}) \Vert^{2}$  

(ii) \textit{Refinement phase}: In the second stage, we iteratively refine the gain and frequency estimates. Suppose that after the $n$th round of optimization, the gain and frequency estimates are given by $\hat{g}_{n}$ and $\hat{\omega}_{n}$ respectively (starting off with the estimates from the detection phase). In the $n+1$th round, we refine the frequency estimate by fixing the gain to $\hat{g}_{n}$ and locally optimizing $S(\hat{g}_{n},\omega)$ around $\hat{\omega}_{n}$ using Newton's method: 
\begin{equation}
\hat{\omega}_{n+1}=\hat{\omega}_n-\frac{\partial S(\hat{g}_{n},\hat{\omega}_n)/\partial \omega}{\partial^2 S(\hat{g}_{n},\hat{\omega}_n)/\partial \omega^2},
\end{equation}
where 
\begin{align}
\frac{\partial S(g,\omega)}{\partial \omega}&=\Re\left\{\left(\mathbf{y}-g\bB\mathbf{x}(\omega)\right)^H g \bB\left({d \mathbf{x}(\omega)}/{d \omega}\right)\right\},\\
\frac{\partial^2 S(g,\omega)}{\partial \omega^2}&=\Re\left\{\left(\mathbf{y}-g\bB\mathbf{x}(\omega)\right)^H g\bB\left({d^2 \mathbf{x}(\omega)}/{d\omega^2}\right)\right\}\nonumber\\
& \ \ \ \ \ \ \ \ \ \ \ \ \ \ \  - |g|^2\left\Vert\bB\left({d \mathbf{x}(\omega)}/{d \omega}\right)\right\Vert^2.
\end{align}
Next, fixing the frequency estimate to $\hat{\omega}_{n+1}$, we get the updated gain after the $n+1$th round to be $\hat{g}_{n+1} = \left(\bB\mathbf{x}(\hat{\omega}_{n+1}))^H\mathbf{y}\right)/\Vert \bB\mathbf{x}(\hat{\omega}_{n+1})\Vert^2$. Our numerical results are based on applying three such rounds of iterative optimization.

\noindent \textbf{Results:} We simulate the performance of the algorithm with $M=10,25,40,60$ and $256$ compressive measurements across {\it effective per measurement SNRs} $M/(N\sigma^2)$ ranging from $-30$dB to $1$dB using $5\times10^4$ trials (for each $M$, we use the same measurement matrix $\bA$ for all SNR values). The elements of $\bA$ are picked i.i.d from $\text{Uniform}\{\pm 1/\sqrt{N}, \pm j/\sqrt{N}\}$. We plot the Root-Mean-Squared-Error (RMSE) of the frequency estimate versus the effective SNR $M/(N\sigma^2)$ along with the CRB and ZZB in Fig.~\ref{fig:RMSE}. We define the effective SNR beyond which the RMSE of the estimate exhibits a linear falloff with SNR in the log-log plot (similar to the ZZB at high SNRs) as the \textit{RMSE threshold}. From our earlier discussions on the number of measurements needed for successful compressive estimation, we expect the RMSE threshold to exceed the ZZB threshold. From Fig.~\ref{fig:RMSE}, we see that the RMSE thresholds for $M=10,25,40,60$ and $256$ measurements are $-2,-6,-7,-7$ and $-8$dB respectively. All the RMSE thresholds are larger than the ZZB threshold of $-10$dB as expected. We also evaluate the algorithm for the all $N$ measurements case ($\bB=\mathbb{I}_N$) and find that the RMSE threshold in this case is $-8$dB.

Differences in the isometry constant $\epsilon$ explain why the RMSE thresholds are different for different measurement matrices $\bA$.  With increasing number of measurements $M$, the isometry constant decreases. This trend is shown in Fig.~\ref{fig:iso_Consts} where we plot the bounds on the deviation of the pairwise SNRs from $M/N$, corresponding to $(1\pm\epsilon)^2$, for the measurement matrices $\bA$ used in our simulations. $\big(${\it Note}: These isometry constants correspond to the manifold $\{ge^{j\phi}\bx(\omega) : g\in\mathbb{R}^{+},\phi,\omega\in[0,2\pi)\}$ because the algorithm does not use the fact that $g=1\big)$. When we take few compressive measurements, pairwise SNRs can deteriorate significantly ($\epsilon$ is large) and, as a result, the RMSE threshold increases. The bounds on pairwise SNR variation (in Fig.~\ref{fig:iso_Consts}) when we make $40$ and $60$ measurements do not differ by much. This illustrates the diminishing improvements in isometry per measurement beyond a point. For the all-$N$ measurements case ($\bB = \mathbb{I}_N$), the isometry constant $\epsilon = 0$ by definition and therefore the RMSE threshold is close to the ZZB threshold.

When we set $M=N=256$, the degradation in pairwise SNRs is smaller than $2$dB. However, for this extreme case, the RMSE threshold is merely $1$dB smaller than that for $M=40$. This indicates that, for our example of frequency estimation for sinusoid of length $N=256$, the isometry constant is small enough 
when we make $40$ or more compressive measurements.

To summarize, when the number of measurements $M$ is large enough for the isometry constant $\epsilon$ to be small, the number of measurements $M$ necessary
obeys the rule of thumb in (\ref{eq:number_measurements_zzb}), based on ZZB threshold computations for the original system.  For our example $N=256$ sinusoid,
this translates to the rule of thumb $M \geq {\rm max} \{ 40, 25.6 \sigma^2 \}$.

\negspace
\section{Isometry conditions for frequency estimation from compressive measurements}
\label{sec:multipleSinusoids}
In the single sinusoid example in the previous section, we assume that there are enough measurements to guarantee the required isometries. In this section, we seek to analytically characterize the number of measurements required to provide such guarantees.  We show that, for a mixture of $K$ sinusoids, the number of measurements required depends on the conditioning of appropriately defined matrices, which in turn depends on the separation between the frequencies in the mixture. We return to the special case of a single sinusoid, for which we can prove stronger results, at the end of this section

Consider a manifold of signals which are linear combinations of $K$ complex sinusoids $\sum_{l=1}^{K} g_{l} \bx(\omega_{l})$, where $g_{l}$ $\in$ $\mathbb{C}$ are complex gains and 
\begin{equation}
\mathbf{x}(\omega) = \left[h_1 e^{-j\omega(N-1)/2}\ \cdots\ h_{N}e^{j\omega(N-1)/2} \right]^T
\label{eq:windowedSinusoid}
\end{equation} 
is a windowed sinusoid, with window weights given by $\{h_{n}\}$. The example in the previous section is a special case with $K=1$ and
an all-ones window. Without loss of generality, we assume that the window weights are normalized so that $\sum_{n} \left| h_{n} \right|^{2} = 1$. To avoid trivialities, we assume that more than one of the $h_{n}$'s are non-zero.

Suppose that we make $M$ compressive measurements of the form
\begin{equation}
\textstyle
\mathbf{y} =  \bm{A}  \sum_{l=1}^{l=K}g_l \mathbf{x}(\omega_l) + \mathbf{z}, 
\label{eq:multiSinusoidMModel}
\end{equation}
and we wish to estimate the gains $\bg=[g_1\ \cdots\ g_K]^T$ and the frequencies $\bomega =[\omega_1\ \cdots\ \omega_K]^T$. Therefore, in the notation of the preceding sections the parameter to be estimated is $\btheta = (\bg,\bomega)$.

\noindent \textbf{Tangent plane isometry for a mixture of $K$ sinusoids:} Our first goal is to quantify the number of measurements needed to preserve the CRB for a given frequency support $\bomega$ (i.e., for all $\btheta$ that share this frequency support). We show that this is equivalent to guaranteeing $\epsilon$-isometry for a {\it set} of tangent planes as follows. For any specific value of the unknown parameters -- gain magnitude $\{|g_{l}|\}$, phases $\{g_{l}/|g_{l}|\}$ and frequencies $\{\omega_{l}\}$ (we split the complex gain in this manner in order to restrict attention to real parameters) -- Theorem \ref{thm:CRB:relation} guarantees that the CRB can be preserved (up to the $M/N$ SNR penalty) by ensuring $\epsilon$-isometry for the plane tangent to the manifold at this set of parameters. Therefore, to preserve the CRB for the frequency support $\bomega$, we need to guarantee $\epsilon$-isometry for tangent-planes for all values that the gain magnitudes $\{|g_{l}|\}$ and the phases $\{g_{l}/|g_{l}|\}$  can take. We can show that the union of all such tangent planes is a subset of the span of the matrix $\bT(\bomega)$ (in $\mathbb{C}^N$), defined as
\begin{equation}
\mathbf{T}(\bm{\omega})\!=\!\left[\mathbf{x}(\omega_1)\ \cdots\ \mathbf{x}(\omega_K)\ \tau\frac{d\mathbf{x}(\omega_1)}{d\omega}\ \cdots\ \tau\frac{d\mathbf{x}(\omega_K)}{d\omega}\right]
\label{eq:tgtPlane}
\end{equation}
where $\tau=1/\Vert d\mathbf{x}(\omega)/d\omega\Vert$ (note that $\tau$ does not depend on $\omega$). Therefore, if the compressive measurement matrix $\bA$ satisfies 
\begin{equation}
\sqrt{\frac{M}{N}}(1-\epsilon) \leq \frac{\Vert \bA \bT(\bomega) \bq \Vert}{\Vert \bT(\bomega) \bq \Vert} \leq \sqrt{\frac{M}{N}} (1 + \epsilon) ~ \forall \bq\in\mathbb{C}^{2K},
\end{equation}
we can preserve the CRB (up to the SNR penalty) for a given frequency support $\bomega$. Furthermore, if the above relationship holds, we say that $\bA$ satisfies the tangent plane $\epsilon$-isometry property at $\bomega$. 

Our first result is to show that the smallest singular value of the matrix $\bT(\bomega)$, given by $\delta = \min_{\bq\in\mathbb{C}^{2K}} \Vert \bT(\bomega) \bq \Vert/\Vert \bq \Vert$, compactly characterizes the number of measurements needed to preserve tangent plane $\epsilon$-isometry.

\begin{theorem}\label{thm:Isometry_mixture_tangent}
Let $ \bm{A} $ be an $M\times N$ measurement matrix whose entries are drawn i.i.d. from $\mathrm{Uniform}$ $\{\pm1/\sqrt{N},\pm j/\sqrt{N}\}$. Let $\bT(\bomega)$ denote the tangent plane matrix (\ref{eq:tgtPlane}) of sinusoids (\ref{eq:windowedSinusoid}) with frequencies $\bomega = (\omega_1 \ldots \omega_K) \in \mathbb{R}^K$.  Let $\Lambda_{T}(\delta) = \{\bomega: \text{smallest singular value of } \bT(\bomega) \geq \delta\}$. Then, for any $\epsilon > 0$, we have
\begin{eqnarray}
	1-\epsilon \leq \sqrt{\frac{N}{M}}\frac{\Vert \bA \bT(\bomega) \bq\Vert}{\Vert \bT(\bomega) \bq \Vert} \leq 1+\epsilon, \forall \bomega \in \Lambda_{T}(\delta),  \bq \in \mathbb{C}^{2K} \label{eq:isometry:mixture_sinusoids_tangent:Thm}
\end{eqnarray}
with high probability when $M=$ $O( \epsilon^{-2} K \log( N K$ $\epsilon^{-1}\delta^{-1} ) )$.  
\end{theorem}
\noindent \textbf{Remarks:}

\noindent $\bullet$ The theorem states that the minimum number of measurements scales as the inverse of the smallest singular value $\delta$.
The singular values of $\bT(\bomega)$ are the square roots of the eigenvalues of $\bT^{H}(\bomega)\bT(\bomega)$,
whose entries can be shown to depend only on the  set of frequency {\it differences} $\omega_{i} - \omega_{j}, 1 \leq i,j \leq K$. Therefore, $\delta$ depends only on the set of frequency differences.

\noindent $\bullet$ The smallest singular value $\delta$ tends to zero when any two of the $K$ frequencies (say $\omega_{i}$ and $\omega_{j}$) get close, since the columns $\bx(\omega_{i})$ and $\bx(\omega_{j})$ (and hence the columns $d\bx(\omega_{i})/d\omega$ and $d\bx(\omega_{j})/d\omega$) approach each other, and the matrix $\bT(\bomega)$ becomes poorly conditioned. It is a natural question, therefore, to ask  whether it is possible to provide a lower bound on $\delta$, and hence an upper bound on the number of measurements required to give tangent plane isometries, by ensuring that the spacing between the constituent frequencies is large enough (larger than say $\Delta\omega$). We leave this as a topic for further investigation, since that characterization of the smallest singular value $\delta$ in terms of the minimum frequency separation $\Delta\omega$ is a feature of the original system with a full set of measurements rather than a problem inherent to compressive estimation. It is interesting to note that prior work on non-compressive frequency estimation \cite{ Recht_2013_with_Noise_no_CS, Recht_2012_denoising, candes:superresolution}, while not directly working with the parameter $\delta$, also requires a minimum frequency separation for successful estimation (e.g., a separation of around four times the DFT spacing of $2 \pi/N$) using $N$ measurements.

 \noindent $\bullet$ When the frequency support $\bomega$ is ``\textit{roughly}" known ahead of time (say $\bomega\approx\bomega_0$), such as in tracking scenarios encountered in radar (where frequencies correspond to directions of arrival), $\bA$ need \textit{only} preserve the norms of vectors in the span of $\bT\left(\bomega\right)$ for $\bomega = \bomega_0$ (not all $\bomega\in\Lambda_{T}(\delta)$). Typically, the number of sinusoids $K$ in the mixture is small. So, one can do better than the $M/N$ SNR penalty that would be incurred if a compressive measurement matrix is used: In such a scenario, it may even be possible to preserve the CRB with no SNR degradation whatsoever. The equivalent problem of direction-of-arrival estimation is studied in \cite{Weiss_1994}. The precise conditions on $\bA$ so that the CRB is preserved with no SNR penalty are stated in \cite{Weiss_1994}. This, however, requires knowing the very frequencies that we wish to estimate. Of course, this is not applicable to the one-shot estimation problem considered here, where we wish to preserve the CRB (up to the SNR penalty $M/N$) with a few measurements $M$, irrespective of what the particular realization of $\bomega$ is.

\noindent \textbf{Pairwise isometry for a mixture of $K$ sinusoids:} Consider now the problem of quantifying the number of measurements needed to guarantee pairwise $\epsilon$-isometry for a mixture of $K$ sinusoids.  We denote the matrix containing the sinusoids $\left[\bx(\omega_1) ~\bx(\omega_2)~\ldots \bx(\omega_K)\right]$ by $\bX(\bomega)$. From the definition of pairwise isometry in Section \ref{sec:modelIsometries}, compressive measurements must preserve the ML cost structure, thereby implying that
\begin{equation}
\left \Vert \bA \bX(\bomega) \bg - \bA \bX(\bomega') \bg' \right \Vert \approx \sqrt{{M}/{N}} \Vert \bX(\bomega) \bg - \bX(\bomega') \bg' \Vert,
\end{equation}
for pairs of $(\bg,\bomega)$ and $(\bg',\bomega')$ of interest. We are typically interested in all values of the gains $\bg,\bg'$ but may restrict the set of frequencies $\bomega$ and $\bomega'$ to each come from a set $\Theta$ (for example, the set of $K$ frequencies that are separated pairwise by at least $\Delta \omega$). 

To simplify the problem, we only consider $\bomega$ and $\bomega'$  that are ``well-separated'' (we comment on why this helps later). For example,  we may restrict $\bomega'$ to $\Theta'(\bomega) = \Theta$\textbackslash$B(\bomega,\mu)$, where $B(\bomega,\mu)$ is a small ball of frequencies around $\bomega$.  (A possible definition for the ball $B(\bomega,\mu)$ can be $B(\bomega,\mu) = \{\bomega^{\prime}:\min_{1\leq i,j\leq K} |\omega_{i}^{\prime} - \omega_j| \leq \mu \}$). Suppose that we make enough measurements to guarantee pairwise $\epsilon$-isometry for all $\bomega \in \Theta$ and $\bomega' \in \Theta'(\bomega)$, no matter what value $\bomega$ takes. This implies that for any set of frequencies $\bomega \in \Theta$, we have preserved the cost-structure of the estimation problem at hypothesis frequencies $\bomega'$ that are ``far-away'' ($\bomega'$ outside $B(\bomega,\mu))$. Roughly, a good estimation algorithm should not incur frequency errors larger than $\mu$ at high SNRs.  

We introduce some notation for the following discussion. Let $\bomegatilde = \left[\bomega~\bomega' \right], \bgtilde = \left[\bg~-\bg' \right]$ denote vectors of length $2K$ concatenating the gains and frequencies. Also let $\bX(\bomegatilde) = \left[\bX(\bomega)~\bX(\bomega^\prime)\right]$ denote the $N \times 2K$ matrix containing all the sinusoids. Note that $\bgtilde$ can take any value in $\mathbb{C}^{2K}$ but $\bomegatilde$ has a special structure: its first $K$ entries $\bomega$ must belong to $\Theta$ and its last $K$ entries come from a set $\Theta'(\bomega)$ that depend on the first $K$ values.  As shorthand, we say that $\bomegatilde \in \tilde{\Theta} = \{[\bomega~\bomega']: \bomega \in \Theta, \bomega' \in \Theta'(\bomega)\}$. With this notation, the above pairwise isometry condition for a mixture of $K$ sinusoids, which we desire can be written as 
\begin{equation}
\sqrt{\frac{M}{N}} (1-\epsilon) \leq \frac{\Vert \bA \bX(\bomegatilde) \bgtilde \Vert}{\Vert \bX(\bomegatilde) \bgtilde \Vert} \leq \sqrt{\frac{M}{N}} (1+\epsilon)\ \forall \bgtilde\in\mathbb{C}^{2K},
\label{eq:isometryActual}
\end{equation}
for a particular $\bomegatilde \in \tilde{\Theta}$. If the matrix $\bA$ satisfies this relationship, we say that $\bA$ guarantees $\epsilon$-isometry ({\it just isometry, not pairwise}) for the frequency support $\bomegatilde$ ($2K$ sinusoids). 

Our goal is to quantify the number of measurements necessary for (\ref{eq:isometryActual}) to hold for all $\bomegatilde \in \tilde{\Theta}$. While solving this problem in its entirety is difficult, we can break it down into two subproblems, the first of which we tackle. We explain the solution to this subproblem and then comment on the other.  In analogy with our previous discussion of tangent plane isometry, let $\Lambda_p(\delta)$ denote the set of all frequencies $\bomegatilde$ (chosen from anywhere in $\mathbb{R}^{2K}$, not just $\tilde{\Theta}$) such that the smallest singular value of $\bX(\bomegatilde)$ is at least as large as $\delta$. Suppose that we want $\bA$ to guarantee $\epsilon$-isometry for all $\bomegatilde \in \Lambda_p(\delta)$ (as in (\ref{eq:isometryActual}) except that the set from which $\bomegatilde$ is chosen has changed). We show that $M=O\left( \epsilon^{-2} (2K) \log\left( N (2K) \epsilon^{-1}\delta^{-1} \right) \right)$ measurements suffice to provide such a guarantee with high probability. 

\begin{theorem}\label{thm:Isometry_mixture}
Suppose that $ \bm{A} $ is an $M\times N$ measurement matrix whose entries are drawn i.i.d. from $\mathrm{Uniform}$ $\{\pm1/\sqrt{N},\pm j/\sqrt{N}\}$. Let $\bX(\bomega) = \left[ \bx(\omega_1)~\bx(\omega_2)~\ldots~\bx(\omega_K) \right]$ denote an $N \times K$ matrix of sinusoids (\ref{eq:windowedSinusoid}) with $\bomega = (\omega_1 \ldots \omega_K) \in \mathbb{R}^K$. Let $\Lambda_p(\delta) = \{\bomega: $ smallest singular value of $\bX(\bomega)$ is greater than or equal to $\delta$\}. For any $\epsilon > 0$ and $\delta>0$,  we have
\begin{eqnarray}	
	1-\epsilon \leq\sqrt{\frac{N}{M}} \frac{\left\Vert \bA \bX(\bomega) \bg \right\Vert}{\left\Vert \bX(\bomega) \bg \right\Vert}\leq 1+\epsilon,  
	\forall \bomega \in \Lambda_p(\delta),\bg \in \mathbb{C}^K, \label{eq:isometry:mixture_sinusoids:Thm}
\end{eqnarray}
with high probability when $M=O\left( \epsilon^{-2} K \log\left( N K \epsilon^{-1}\delta^{-1} \right) \right)$.  
\end{theorem}

\noindent \textbf{Remarks:} 

\noindent $\bullet$ Returning to the problem posed in (\ref{eq:isometryActual}), suppose that the smallest singular value of $\bX(\bomegatilde)$, further minimized over all values of $\bomegatilde \in \tilde{\Theta}$ is $\sigma_{min} > 0$. Then, $\tilde{\Theta}$ is contained in $\Lambda_{p}(\sigma_{min})$ and using Theorem \ref{thm:Isometry_mixture}, $M=O\left( \epsilon^{-2} (2K) \log\left( N (2K) \epsilon^{-1}\sigma_{min}^{-1} \right) \right)$ measurements suffice to guarantee the required $\epsilon$-isometry. 

\noindent $\bullet$ While the singular values of $\bX(\bomegatilde)$ depend only on frequency differences, we leave the question of quantifying $\sigma_{min}$ (e.g., in terms of the minimum pairwise separation $\Delta\omega$ of frequencies for $\bomega\in\Theta$) and $\mu$ (the radius of the ball around each $\bomega\in\Theta$) as an open problem. The problem of lower bounding the singular values of the Fourier matrix ($\bX(\bomega)$, when choosing $\{h_n\}$ in (\ref{eq:windowedSinusoid}) as the all-ones sequence) as a function of minimum frequency separation has been investigated in \cite{ferreira_super-resolution_1999} (using Gershgorin-type bounds). Similar ideas may be useful in our present context as well, but again, these are fundamental and difficult questions regarding the original frequency estimation problem (with a full set of measurements) that are beyond our scope here. We are, however, able to provide an explicit characterization for the special case of a single sinusoid in Appendix \ref{appendix:proof:thm:singleSine}. 

\noindent $\bullet$ The previous remark also explains why we choose to restrict $\bomega'$ to $\Theta'(\bomega) = \Theta$\textbackslash$B(\bomega,\mu)$. The singular value of $\bX(\bomegatilde)$ when $\bomega,\bomega' \in \Theta$ can be made arbitrarily small by allowing $\bomega' \rightarrow \bomega$. Thus, in this case, we cannot directly use Theorem \ref{thm:Isometry_mixture} to quantify the number of measurements required. However, this does not necessarily mean that an isometry cannot be provided for closely spaced sinusoids. Indeed, we show in Appendix \ref{appendix:proof:thm:singleSine} that, for $K=1$, it is possible to provide an isometry no matter how close $\omega$ and $\omega'$ get.

\noindent\textit{Proof of Theorems \ref{thm:Isometry_mixture_tangent} and \ref{thm:Isometry_mixture}:}
We give a proof of Theorem \ref{thm:Isometry_mixture} along the lines of the proof in \cite{baraniuk_random_2009}, where the authors extend the JL lemma (which gives the number of compressive measurements needed to preserve the geometry of a discrete point cloud) to a manifold by sampling the manifold and exploiting its continuity. Details of the proof can be found in Appendix \ref{appendix:proof:thm:pairwise}. A similar proof can be given for Theorem \ref{thm:Isometry_mixture_tangent}, which we briefly sketch in Appendix \ref{appendix:proof:thm:tangent}.


\subsection{Pairwise isometry for frequency estimation of a single sinusoid} \label{sub:sec:SingleSinusoid}
In the preceding discussions, we quantify the number of measurements needed to give pairwise isometries for a mixture of $K$ sinusoids in two distinct regimes: when the frequencies $(\bomega,\bomega^{\prime})$ are ``far apart'' and in the limit of $\bomega^{\prime} \rightarrow \bomega$ (tangent plane isometries). We now consider a single sinusoid ($K=1$) and provide pairwise isometries for {\it all} frequency pairs. In order to do this, we consider two regimes of frequency pairs $(\omega_{1},\omega_{2})$: closely spaced and well-separated. For the set of well-separated frequencies, say $\{(\omega_{1},\omega_{2}): |\omega_{1}-\omega_{2}| > \psi\}$, we obtain a bound on the smallest singular value of $\bX(\omega) = [\bx(\omega_{1})~\bx(\omega_{2})]$ and use it in Theorem \ref{thm:Isometry_mixture} to immediately infer the number of measurements needed to guarantee pairwise $\epsilon$-isometry for sinusoids from this set. The challenge then is in providing a similar result for sinusoids whose frequencies are separated by less than $\psi$. We solve this problem in two stages: first, we use Theorem \ref{thm:Isometry_mixture_tangent} to infer the number of measurements needed to guarantee tangent plane $\epsilon$-isometries for all frequencies (loosely, pairwise isometries for $\omega_{1} \rightarrow \omega_{2}$). We then use the continuity of the sinusoidal manifold to extend these tangent plane $\epsilon$-isometries to a pairwise $2\epsilon$-isometry for closely-spaced frequencies $\{(\omega_{1},\omega_{2}): |\omega_{1}-\omega_{2}| < \psi\}$. 

\begin{theorem}
\label {thm:singleSine}
Suppose that $ \bm{A} $ is an $M\times N$ measurement matrix whose entries are drawn i.i.d. from $\mathrm{Uniform}$ $\{\pm1/\sqrt{N},\pm j/\sqrt{N}\}$. Let $\bx(\omega)$ denote a sinusoid (\ref{eq:windowedSinusoid}) of frequency $\omega$ with weights $\{h_n\}$ such that $\sum |h_n|^2 = 1$. Let $H(\omega)=\sum_{n=1}^{n=N} |h_n|^2 e^{j\omega (n-(N+1)/2)}$ be such that (i) the maxima of $|H(\omega)|^2$ that occur at frequencies other than $\omega =0 $ (side-lobes) are smaller than some constant $D<1$ (independent of $N$) and (ii) $|H(\omega)|^2$ is non-increasing in $(0,\pi/(2N))$. Then, for any $\epsilon > 0$, 
\begin{equation}
1-\epsilon \leq \sqrt{\frac{N}{M}} \frac{\Vert g_{1} \bA \bx(\omega_{1}) - g_{2} \bA \bx(\omega_{2}) \Vert}{\Vert g_{1}\bx(\omega_{1}) - g_{2} \bx(\omega_{2}) \Vert} \leq 1 + \epsilon, ~ \forall g_{1},g_{2},\omega_{1},\omega_{2}
\end{equation}
with high probability when $M=O( \epsilon^{-2}\log( N \epsilon^{-1} (1-\tau \chi)^{-1} \zeta^{-1} \alpha^{-1} ))$ where $\tau = 1/\Vert d\bx(\omega)/d\omega \Vert$, $\chi = \left|dH(0)/d\omega\right|$, $\alpha = 1/(N\tau)$ and $\zeta = -\frac{N^{-2}}{2} \frac{d^{2}|H(0)|^{2}}{d\omega^{2}}$ are parameters of the windowing sequence $\{|h_n|^2\}$.
\end{theorem}
We give the proof of Theorem \ref{thm:singleSine} in Appendix \ref{appendix:proof:thm:singleSine}. The condition that (i) $|H(\omega)|^2$ is monotonic in $(0,\pi/(2N))$ and (ii) all side-lobes peaks of $|H(\omega)|^2$ are smaller than an absolute constant $D<1$ (the main-lobe peak $|H(0)|^2= 1$ since $\sum|h_n|^2 = 1$) are mild. These conditions are satisfied by windowing sequences $\{|h_n|^2\}$ commonly used for spectral estimation, such as the all-ones, Hamming, Hanning, Triangular and Blackman sequences.

\noindent \textbf{Remark:} We state and prove Theorems \ref{thm:Isometry_mixture_tangent}, \ref{thm:Isometry_mixture} and \ref{thm:singleSine} for compressive measurements with projection weights (elements of the matrix $\bA$) taken from $X\sim \text{Uniform}\{\pm 1/\sqrt{N},\pm j/\sqrt{N}\}$. In addition to the concentration results on $\Vert\bA\bv\Vert^2$ of the form (\ref{eq:concentration:projection}) which we need to preserve the geometry of a discrete point cloud, we use the fact that the Frobenius norm $\Vert\bA\Vert_F = \sqrt{M}$ w.p. 1 for this choice of distribution $X$. When the elements of $\bA$ are drawn from other distributions such as the gaussian distribution for which these concentration results on $\Vert\bA\bv\Vert^2$ are also available \cite{vershynin_introduction_2010}, $\Vert\bA\Vert_F^2$, which is the sum of the square of all elements of $\bA$, can be shown to fall within $M(1\pm\delta)$ w.h.p. Therefore, the conclusions of Theorems \ref{thm:Isometry_mixture_tangent},  \ref{thm:Isometry_mixture} and \ref{thm:singleSine} also apply when the elements of $\bA$ are drawn from these distributions.
\negspace
\section{Conclusions and Future Work}
\label{sec:conclusions}

For parameter estimation in AWGN, we have identified isometry conditions under which the only effect of making compressive measurements is an SNR penalty equal to the dimensionality reduction factor. We prove this by establishing a connection between the isometry conditions and the CRB/ZZB. For a mixture of $K$ sinusoids of length $N$, we show that $O(K \log NK \delta^{-1})$ measurements suffice to provide such isometries, where $\delta$ is the smallest singular value of appropriate matrices (stronger results are obtained for $K=1$). Based on the connection between the ZZB and CRB, we also observe that, in order to avoid large estimation errors, the compressive measurements must not only preserve the geometry, but the SNR after the dimension reduction penalty must also be above a threshold.  We illustrate this by showing that, for frequency estimation for a single sinusoid, the convergence of the ZZB to the CRB can be used to tightly predict the number of measurements needed to avoid error floors.

We leave open the issue of establishing the relationship between the smallest singular value $\delta$ and the minimum spacing between sinusoids, and whether the stronger isometry results established for a single sinusoid can be extended to $K>1$. Another interesting topic for future work is the development of an analytical understanding of multi-dimensional sinusoid estimation, motivated by practical applications such as large 2D arrays for mm-wave communication \cite{allerton:tracking} and imaging. Finally, investigation of compressive parameter estimation in non-Gaussian settings is an interesting problem with few known results.

\negspace
\appendices

\section{(Extended) Ziv-Zakai Bound Review\cite{Bell::EZZB::1997}}
\label{appendix:zzbReview}
Consider the problem of estimating a parameter $\btheta$ from measurements 
\begin{equation}
\by =\bs(\btheta) + \bz,~\btheta \in \Theta, \bz \sim \mathcal{CN}(\mathbf{0},\sigma^{2} \mathbb{I}).
\end{equation}
For an estimator $\hat{\btheta}(\by)$, let $\bepsilon = \hat{\btheta}(\by) - \btheta$ denote the estimation error.
The ZZB lower bounds the error $\mathbb{E}|\ba^T\bepsilon|^2$ for any $\ba\in\mathbb{R}^{K}$ by relating it to the probabilities of error in a sequence of detection problems. We begin by describing one of the detection problems.

Consider a simplified version of the preceding model, in which the parameter $\btheta$ takes only two values $\bphi$ and $\bphi + \bdelta$, occurring with probabilities $p(\bphi)/\left(p(\bphi) + p(\bphi + \bdelta) \right)$ and $p(\bphi + \bdelta)/\left(p(\bphi) + p(\bphi + \bdelta) \right)$, respectively. There are two possible ways to estimate $\btheta$:

\noindent $\bullet$ {\it Optimal detection-theoretic approach:} Compute the Bayesian posterior probabilities $p(\bphi|\by)$ and $p(\bphi+\bdelta|\by)$. Choose $\bphi$ if $p(\bphi|\by) > p(\bphi +\bdelta|\by)$ and $\bphi + \bdelta$ otherwise. Denote the probability of error with this approach by $f(\bphi,\bphi + \bdelta)$.

\noindent $\bullet$ {\it Heuristic approach using the estimate $\hat{\btheta}(\by)$:} Form the estimate $\hat{\btheta}(\by)$; this could take any value in $\Theta$,
and is not restricted to $\{\bphi,\bphi + \bdelta\}$. Classify based on the following rule: if $\ba^{T} \hat{\btheta}(\by) < \ba^{T} \bphi + (h/2)$, where $h = \ba^{T}\bdelta$, choose $\bphi$ to have occurred; else, choose $\bphi + \bdelta$. Denote the probability of error with this scheme by $P_{sub}(\bphi,\bphi + \bdelta)$.

Since the Bayesian detection rule is optimal, we have $f(\bphi,\bphi + \bdelta) \leq P_{sub}(\bphi,\bphi + \bdelta)$.

In order to use this observation to bound $\mathbb{E}|\ba^T\bepsilon|^2$, we begin with the identity
\begin{equation}
\mathbb{E}\left|\ba^T\bepsilon\right|^2 = \frac{1}{2}\int_0^\infty \Pr\left(\left|\ba^T\bepsilon\right|\geq h/2\right) h\ dh,
\label{eq:relateEstDetect}
\end{equation}
and relate $\Pr\left(\left|\ba^T\bepsilon\right|\geq h/2\right)$ to the probability of error with the heuristic rule $P_{sub}\left(\bphi,\bphi+\bdelta\right)$ as follows:
\begin{multline}
\Pr\left(\left|\ba^T\bepsilon\right|\geq h/2\right) = \int_{\mathbb{R}^K}\left(p(\bphi) + p(\bphi + \bdelta) \right)\\ P_{sub}\left(\bphi,\bphi+\bdelta \right)\ d\bphi,
\label{eq:relateTailPToPSub}
\end{multline}
where $\bdelta$ is {\it any} vector satisfying $\ba^{T}\bdelta = h$. We now use the lower bound $P_{sub}\left(\bphi,\bphi+\bdelta \right) \geq f(\bphi,\bphi + \bdelta)$ in (\ref{eq:relateTailPToPSub}) and substituting back in (\ref{eq:relateEstDetect}), we get the basic version of the ZZB. 

We can further tighten the bound in two ways: (a) by choosing $\bdelta$ appropriately and (b) by exploiting the fact that $\Pr\left(\left|\ba^T\bepsilon\right|\geq h/2\right)$ is non-increasing using the valley filling operation $\mathcal{V}\{\ \}$, defined as $\mathcal{V}\{q(h)\}=\max_{r \geq 0}q(h+r)$ (refer \cite{Bell::EZZB::1997} for details). This gives us the ZZB in (\ref{eq:ZZB:bound}).

\negspace
\section{Proof of Theorem \ref{thm:Isometry_mixture}}\label{appendix:proof:thm:pairwise}

Let $\bm{\omega}=[\omega_1\ \cdots\ \omega_K]^T$, $\mathbf{g}=[g_1\ \cdots\ g_K]^T$ and $\mathbf{X}(\bm{\omega}) = [\mathbf{x}(\omega_1)\ \cdots\ \mathbf{x}(\omega_K)]$. We note that an $\epsilon$-isometry for all vectors of the form $\mathbf{X}(\bm{\omega})\mathbf{g}$ such that $\Vert\mathbf{X}(\bm{\omega})\mathbf{g}\Vert>\delta$ and $\Vert\mathbf{g}\Vert=1$ is equivalent to (\ref{eq:isometry:mixture_sinusoids:Thm}). We discretize the frequencies $[0,2\pi]$ uniformly into $R$ points ($R$ is specified later) and obtain the set $F$. We first prove a $2 \epsilon_0$ isometry for all vectors in the span of $\mathbf{X}(\mathbf{q})$ for all frequency tuplets $\mathbf{q} \in F^K$ (i.e., $\mathbf{q}=[q_1\ \cdots\ q_K]^T$ with $q_l\in F$). We then extend this to a $3.5 \epsilon_0$ isometry for vectors $\mathbf{X}(\bm{\omega})\mathbf{g}$ such that $\Vert\mathbf{X}(\bm{\omega})\mathbf{g}\Vert>\delta$ and $\Vert\mathbf{g}\Vert=1$ by: (a) approximating them to nearby points in the span of $\mathbf{X}(\mathbf{q})$, (b) choosing $R = O(N^{1.5}K^{0.5} \delta^{-1}\epsilon_0^{-1})$ so that the approximation is good.

\noindent \textbf{Sampling:} For any tuplet of sampled frequencies $\mathbf{q} \in F^K$, if $ \bm{A} $ preserves the norm of $\left(6\epsilon_0^{-1}\right)^{2K}$ well-chosen samples in the span of $\mathbf{X}(\mathbf{q})$ up to $\epsilon_0<2/5$, it can be shown that $ \bm{A} $ will preserve the norms of all vectors in the span of $\mathbf{X}(\mathbf{q})$ up to $2 \epsilon_0$ \cite{baraniuk_simple_RIP_2008} (since we are concerned only with $\ell_2$ distances from sampled points, we map the unit ball in $\mathbb{C}^K$ to the unit ball in $\mathbb{R}^{2K}$ using the map $f(\bz) = [\Re\{\bz^T\}~\Im\{\bz^T\}]^T$ and use corresponding covering arguments in \cite{baraniuk_simple_RIP_2008}. The other argument used in \cite{baraniuk_simple_RIP_2008}, which is closure w.r.t. to addition is satisfied in $\mathbb{C}^K$ as well). Since there are $R^K$ sampled frequency tuplets $\mathbf{q}\in F^K$, by demanding that $ \bm{A} $ preserves the norm of $R^K\left( 6\epsilon_0^{-1} \right)^{2K}$ samples, we can provide a $2 \epsilon_0$ isometry for the span of $\mathbf{X}(\mathbf{q}) ~\forall \mathbf{q} \in F^K$.

\noindent \textbf{Isometry for mixtures of arbitrary frequencies:} We now extend this to an $3.5 \epsilon_0$ isometry result for vectors of the form $\mathbf{X}(\bm{\omega})\mathbf{g}$ such that $\Vert\mathbf{X}(\bm{\omega})\mathbf{g}\Vert>\delta$ and $\Vert\mathbf{g}\Vert=1$ by choosing $R$ appropriately.

Let $\mathbf{q}$ be a tuplet in $F^K$ that is close to $\bm{\omega}$ satisfying $\max_l \left|q_l-\omega_l\right| \leq \pi /R$. We let $\mathbf{e}_l=\mathbf{x}(\omega_l)-\mathbf{x}(q_l)$ and bound the absolute value of each term of $\mathbf{e}_l$ using the mean value theorem to get $\left\Vert \mathbf{e}_l\right\Vert \leq   \pi N/(\sqrt{2}R)$.  We use this to calculate a bound on the difference between a vector $\mathbf{X}(\bm{\omega}) \mathbf{g}$ and its approximated version $\mathbf{X}(\mathbf{q}) \mathbf{g}$. Using the definition of $\mathbf{e}_l$, we obtain
\begin{equation}\label{eq:approximation:errors}
	\mathbf{X}(\bm{\omega})\mathbf{g}=\mathbf{X}(\mathbf{q})\mathbf{g}+\sum_{l=1}^{l=K}g_l\mathbf{e}_l.
\end{equation}
Using the triangle inequality and the fact that $\sum_l |g_l|\leq \sqrt{2K}$ (since $\Vert\mathbf{g}\Vert=1$), we have
\begin{equation}
\frac{\left\Vert\mathbf{X}(\mathbf{q})\mathbf{g}\right\Vert}{\left\Vert\mathbf{X}(\bm{\omega})\mathbf{g}\right\Vert} ~ \sphericalangle ~ 1 \pm \frac{\pi \sqrt{K}N R^{-1}}{\left\Vert\mathbf{X}(\bm{\omega})\mathbf{g}\right\Vert}.
\label{eq:bounds:norms}
\end{equation}
where $x~ \sphericalangle ~y \pm z$ denotes $y-z \leq x \leq y+z$.

Next, we bound the difference between the vectors $ \bm{A}  \mathbf{X}(\bm{\omega}) \mathbf{g}$ and $  \bm{A}  \mathbf{X}(\mathbf{q}) \mathbf{g}$. We see that $\Vert \bm{A} \Vert_F=\sqrt{M}$ and, therefore, have $\left\Vert \bm{A}  \mathbf{e}_k\right\Vert \leq  \sqrt{M} \left\Vert\mathbf{e}_k\right\Vert$. Furthermore, since $ \bm{A} $ preserves the norms of all vectors of the form $\mathbf{X}(\mathbf{q})\mathbf{g}$, where $\mathbf{q} \in F^K$ up to an isometry constant $2\epsilon_0$ (and scale factor of $\sqrt{M/N}$), we get
\begin{equation}
     \sqrt{\frac{N}{M}}\frac{\left\Vert \bm{A} \mathbf{X}(\bm{\omega})\mathbf{g}\right\Vert}{\left\Vert\mathbf{X}(\mathbf{q})\mathbf{g}\right\Vert} ~ \sphericalangle ~1 
  \pm \left(2\epsilon_0+ \frac{\pi N\sqrt{N K}R^{-1}}{\left\Vert\mathbf{X}(\mathbf{q})\mathbf{g}\right\Vert}\right). 
     \label{eq:mapping:bounds:norms}
\end{equation}

Before we proceed to give the isometry result, we need to characterize how small $\left\Vert\mathbf{X}(\mathbf{q})\mathbf{g}\right\Vert$ can be in (\ref{eq:mapping:bounds:norms}). Since $\Vert\mathbf{X}(\bm{\omega})\mathbf{g}\Vert>\delta$, from (\ref{eq:bounds:norms}) we have the following:
\begin{equation}
\frac{\left\Vert\mathbf{X}(\mathbf{q})\mathbf{g}\right\Vert}{\left\Vert\mathbf{X}(\bm{\omega})\mathbf{g}\right\Vert} ~ \sphericalangle ~ 1 \pm \pi \sqrt{K}N(R\delta)^{-1}.
\end{equation}
Choosing $R=(4\pi) N\sqrt{NK}\epsilon_0^{-1}\delta^{-1}$, we have that 
\begin{equation}
	\frac{\left\Vert\mathbf{X}(\mathbf{q})\mathbf{g}\right\Vert}{\left\Vert\mathbf{X}(\bm{\omega})\mathbf{g}\right\Vert} ~ \sphericalangle ~ 1 \pm 0.25\epsilon_0.
	\label{eq:relate:to:grid} 
\end{equation}
For this choice of $R$, from (\ref{eq:mapping:bounds:norms}), we see that
\begin{equation}
     \sqrt{\frac{N}{M}}\frac{\left\Vert \bm{A} \mathbf{X}(\bm{\omega})\mathbf{g}\right\Vert}{\left\Vert\mathbf{X}(\mathbf{q})\mathbf{g}\right\Vert} ~ \sphericalangle ~1 
  \pm \left(2\epsilon_0+ \frac{0.25\epsilon_0\delta}{\left\Vert\mathbf{X}(\mathbf{q})\mathbf{g}\right\Vert}\right). 
\end{equation}
Using the lower bound from (\ref{eq:relate:to:grid}), $\left\Vert\mathbf{X}(\mathbf{q})\mathbf{g}\right\Vert\geq(1-0.25\epsilon_0)\left\Vert\mathbf{X}(\bm{\omega})\mathbf{g}\right\Vert\geq\delta(1-0.25\epsilon_0)$,
\begin{equation}
     \sqrt{\frac{N}{M}}\frac{\left\Vert \bm{A} \mathbf{X}(\bm{\omega})\mathbf{g}\right\Vert}{\left\Vert\mathbf{X}(\mathbf{q})\mathbf{g}\right\Vert} ~ \sphericalangle ~1 
  \pm 2.5\epsilon_0. 
\end{equation}
Substituting the bounds for $\left\Vert\mathbf{X}(\mathbf{q})\mathbf{g}\right\Vert$ in terms of $\left\Vert\mathbf{X}(\bm{\omega})\mathbf{g}\right\Vert$ from (\ref{eq:relate:to:grid}), we have that
\begin{equation}
\sqrt{\frac{N}{M}} \frac{\left\Vert\bm{A}\mathbf{X}(\bm{\omega})\mathbf{g}\right\Vert}{\left\Vert\mathbf{X}(\bm{\omega})\mathbf{g}\right\Vert} ~ \sphericalangle ~ 1 \pm 3.5\epsilon_0.
\end{equation}

\noindent\textbf{Number of measurements:} It only remains to specify the number of measurements $M$ required to preserve the norms of the $R^K\left( 6\epsilon_0^{-1} \right)^{2K}$ samples up to $\epsilon_0$. Using the value for $R$ just obtained, and setting $\epsilon = 3.5 \epsilon_0$, we see that we must preserve the norms of $(18\times7^3\pi N^{1.5}K^{0.5}\epsilon^{-3}\delta^{-1})^K$ vectors (samples) up to $2\epsilon/7$ w.h.p. We relate the probability of preserving these norms to the number of measurements $M$ via the concentration results (\ref{eq:concentration:projection}) for $\text{Uniform}\{\pm 1/\sqrt{N}, \pm j/\sqrt{N}\}$ (setting $\delta$ in (\ref{eq:concentration:projection}) and (\ref{eq:concentration:rate}) to $32\epsilon/49$ -- here we have used the fact that when $\epsilon<1$, $\max \{(1+2\epsilon/7)^2-1,1-(1-2\epsilon/7)^2\} < 32\epsilon/49$). We employ the union bound and (\ref{eq:concentration:projection}) to compute the probability that the norm of atleast one sample is not preserved. This probability becomes vanishingly small for $M=O\left(\epsilon^{-2}K\log\left(NK\epsilon^{-1}\delta^{-1}\right)\right)$ measurements, which concludes the proof.

\negspace

\section{Proof of Theorem \ref{thm:Isometry_mixture_tangent}}\label{appendix:proof:thm:tangent}
For the matrix $\mathbf{T}(\bomega)$,
$K$ of the columns are of the form $\tau d\bx(\omega)/d\omega$, while the remaining $K$ are of the form $\bx(\omega)$. 
When $\tau d\bx(\omega)/d\omega$ is approximated by $\tau d\bx(q)/d\omega$, where $q$ is the frequency on an uniformly spaced frequency grid with $R$ points that is the closest to $\omega$, the norm of the approximation error is upper bounded by $\pi N/(\sqrt{2}R)$. The upper bound on the norm of the error in approximating $\bx(\omega)$ by $\bx(q)$ used in theorem \ref{thm:Isometry_mixture} is also $\pi N/(\sqrt{2}R)$. Therefore, by following the proof of theorem \ref{thm:Isometry_mixture} with $K$ set to $2K$ (because number of columns of $\mathbf{X}(\bomega)$ is only $K$), we obtain the proof for theorem \ref{thm:Isometry_mixture_tangent}.

\negspace

\section{Proof of Theorem \ref{thm:singleSine}}
\label{appendix:proof:thm:singleSine}
We present the results for closely spaced frequencies first (tangent plane isometries), and then move to the well-separated setting.

\noindent \textbf{Tangent plane isometry:} For a single sinusoid, the tangent plane matrix at $\omega$ is given by $\bT(\omega) = \left[\bx(\omega)~ \tau d\bx(\omega)/d\omega \right]$ where $\tau = 1/\Vert d\bx(\omega)/d\omega \Vert$. The smallest singular value of $\bT(\omega)$, denoted by $\sigma_{\textrm{tangent}}$, satisfies 
\begin{eqnarray}
\sigma_{\textrm{tangent}}^2 &=&1-\tau\left|\left<\mathbf{x}(\omega),{d\mathbf{x}(\omega)}/{d\omega}\right>\right| \\ 
&=& 1 - \tau \left| \sum_{n=1}^{n=N} |h_{n}|^{2} j\omega \left(n-({N+1}/{2}) \right) \right| 
\end{eqnarray}
where the second equality is obtained from the definition of the sinusoid (\ref{eq:windowedSinusoid}) by noting that the $n$th entry of $\bx(\omega)$ is $h_{n}e^{j\omega (n-(N+1)/2)}$.
From the definition of $H(\omega)$, we see that,
\begin{equation}
\sigma_{\textrm{tangent}}^2 = 1 - \tau \left|{d H(0)}/{d \omega}\right|, 
\end{equation}
and therefore $\sigma_{\textrm{tangent}}=\sqrt{1-\tau\chi}$ where $\chi=\left|d H(0)/d\omega \right|$. By Jensen's inequality, we see that $\chi^2<1/\tau^2$  when the weight sequence $\{h_n\}$ has more than one non-zero tap. Thus, $\tau \chi < 1$ and therefore $\sigma_{\textrm{tangent}}$ is strictly positive. Setting $\delta = \sqrt{1 - \tau \chi}$ in Theorem \ref{thm:Isometry_mixture_tangent}, we can provide tangent plane $\epsilon$-isometries for a single sinusoid with $M=O\left( \epsilon^{-2}  \log\left( N  \epsilon^{-1}(1-\tau\chi)^{-1} \right) \right)$ measurements.

\noindent \textbf{Extending tangent plane isometry to pairwise isometry for frequencies separated by at most $1/N^{1.5}$:} We now extend $\epsilon$-isometry of the tangent planes to a pairwise $2\epsilon$-isometry for any two frequencies $\omega_{1},\omega_{2}$ whose separation $\Delta = \omega_{2} - \omega_{1}$ is ``small'' (we quantify how small later) by exploiting continuity. Let $q = (\omega_{1}+\omega_{2})/2$ be the average of the two frequencies. For small values of $|\Delta|$, a first-order Taylor series expansion for $\bx(\omega_{1})$ and $\bx(\omega_{2})$ around $\bx(q)$ will have small errors. Such an expansion gives us 
\begin{eqnarray}
\mathbf{x}(\omega_1)&\!\!\! =&\!\!\!\mathbf{x}(q)-(\Delta/2)({d\mathbf{x}(q)}/{d\omega})+\mathbf{e}_1,\\ 
\mathbf{x}(\omega_2)&\!\!\! =&\!\!\!\mathbf{x}(q)+(\Delta/2)({d\mathbf{x}(q)}/{d\omega})+\mathbf{e}_2,
\end{eqnarray}
where $\mathbf{e}_{1},\mathbf{e}_{2}$ are the approximation errors. Consider a linear combination $\bX(\bomega) \bg$ where $\bX(\omega) = \left[\bx(\omega_{1}) ~\bx(\omega_{2}) \right]$ and $\bg = [g_{1}~g_{2}]$. This can be written as 
\begin{equation}
\mathbf{X}(\bm{\omega})\mathbf{g}=\mathbf{v}+\mathbf{e}
\end{equation}
where $\mathbf{e} = g_{1}\be_{1} + g_{2}\be_{2}$ and 
\begin{equation}
\mathbf{v} = \left(g_1+g_2\right)\mathbf{x}(q)
+(\Delta/2)\left(g_2-g_1\right)({d\mathbf{x}(q)}/{d\omega}),
\end{equation}
lies in the span of $\bT(q) = [\mathbf{x}(q)\ \tau(d\mathbf{x}(q)/d\omega)]$, the tangent plane at $\omega = q$. 

Since $\bA$ guarantees $\epsilon$-isometries for tangent planes at all frequencies, for any vector $\bT(q) \bh$ in the tangent plane at $q$, the quantity $\Vert \bA \bT(q) \bh \Vert$ is bounded within $(1\pm \epsilon)\sqrt{M/N} \Vert \bT(q) \bh \Vert$. Expanding out $\Vert \bA \bX(\bomega) \bg \Vert/\Vert \bX(\bomega) \bg \Vert$ in terms of $\bv$ and $\be$ and applying the tangent plane isometry condition to $\Vert \bA \bv \Vert/\Vert \bv \Vert$, we can show that 
\begin{equation}
\sqrt{\frac{N}{M}} \frac{\Vert\bm{A}\mathbf{X}(\bm{\omega})\mathbf{g}\Vert}{\Vert\mathbf{X}(\bm{\omega})\mathbf{g}\Vert} ~ \sphericalangle ~ 1 ~ \pm ~ \left(\epsilon + \frac{5\sqrt{N}\Vert\mathbf{e}\Vert}{\Vert\mathbf{v}\Vert}\right).
\end{equation}
where $x~ \sphericalangle ~y \pm z$ denotes $y-z \leq x \leq y+z$. Next, we get bounds on $\Vert \be \Vert$ and $\Vert \bv \Vert$ as follows. First, we use the mean value theorem to show that the error is bounded as $\Vert \be \Vert \leq N^{2} \Delta^{2}/(4 \sqrt{2})$. Next, since $\bv$ lies in the span of $\bT(q)$, we can use the bound on the minimum singular value of $\bT(q)$ to get $
\Vert\mathbf{v}\Vert\geq \sqrt{1-\tau\chi} |\Delta|/(\sqrt{2}\tau)$. The details are given in Appendix \ref{appendix:extendTgtPlane}. Substituting these bounds in the above equation, we obtain
\begin{equation}
\sqrt{\frac{N}{M}} \frac{\Vert\bm{A}\mathbf{X}(\bm{\omega})\mathbf{g}\Vert}{\Vert\mathbf{X}(\bm{\omega})\mathbf{g}\Vert} ~ \sphericalangle ~ 1 ~ \pm ~ \left(\epsilon + \frac{5\tau|\Delta| N^{2.5}}{4\sqrt{1-\tau\chi}}\right). \label{eqn:def:zeta}
\end{equation}
We note that $\tau = 1/\Vert d \bx(\omega)/d\omega \Vert$ scales as $1/N$. Therefore, defining a scale-invariant constant $\alpha = 1/(N\tau)$, we see that, as long as the frequency separation $|\Delta|\leq (4\alpha\epsilon\sqrt{(1-\tau\chi)}/5)/N^{1.5}$, we can get a $2\epsilon$ isometry
\begin{equation}
\sqrt{\frac{N}{M}} \frac{\Vert\bm{A}\mathbf{X}(\bm{\omega})\mathbf{g}\Vert}{\Vert\mathbf{X}(\bm{\omega})\mathbf{g}\Vert} ~ \sphericalangle ~ 1 ~ \pm ~ 2\epsilon.
\end{equation}

Thus, if $\bA$ provides an $\epsilon/2$ tangent plane isometry for all frequencies (which can be achieved with $M=O\left( \epsilon^{-2}  \log\left( N  \epsilon^{-1}(1-\tau\chi)^{-1} \right) \right)$ measurements), we can extend it to a pairwise $\epsilon$-isometry for the set of frequencies $\omega_{1},\omega_{2}$ whose separation $|\omega_{1}-\omega_{2}| \leq (4\alpha(\epsilon/2)\sqrt{(1-\tau\chi)}/5)/N^{1.5}$.

\noindent \textbf{Pairwise isometry for frequencies separated by more than $1/N^{1.5}$:} We now use Theorem \ref{thm:Isometry_mixture} to quantify the number of measurements necessary to guarantee pairwise $\epsilon$-isometry for two frequencies that are separated by more than $\mu/N^{1.5}$, where $\mu = (4\alpha(\epsilon/2)\sqrt{(1-\tau\chi)}/5)$. 

First, we obtain a bound on the smallest singular value of $\bX(\omega_{1},\omega_{2}) = [\bx(\omega_{1})~\bx(\omega_{2})]$. Denoting the smallest singular value by $\sigma_{\textrm{signal}}^{2}$, we can show that it satisfies
\begin{equation}
\sigma_{\textrm{signal}}^2 =1-\left|\left<\mathbf{x}(\omega_1),\mathbf{x}(\omega_2)\right>\right|.
\end{equation}
Furthermore, we can show that $\left|\left<\mathbf{x}(\omega_1),\mathbf{x}(\omega_2)\right>\right|=|H(\omega_1-\omega_2)|$, where $H(\omega)=\sum_{n=1}^{n=N} |h_n|^2 e^{j\omega (n-(N+1)/2)}$. Thus, we have $\sigma_{\textrm{signal}}^2=1-|H(\omega_1-\omega_2)|$. 

Suppose now that $|\omega_{1}-\omega_{2}| > \mu/N^{1.5}$. For large values of $N$, the smallest singular value of $\bX(\omega_{1},\omega_{2})$ is bounded as
\begin{equation}
\sigma_{\textrm{signal}} > \sqrt{\frac{0.4 \zeta \mu^2}{N}},~\text{where}~ \zeta=-\frac{N^{-2}}{2!}\left.\frac{d^2 |H(\omega)|^2}{d\omega^2} \right|_{\omega=0}\!\!\!. 
\end{equation}
The details are given in Appendix \ref{appendix:singValWellSep}. 

We now apply Theorem \ref{thm:Isometry_mixture} with $\delta = \sqrt{0.4 \zeta \mu^2/N}$. The set of all frequencies $|\omega_{1} - \omega_{2}| > \mu/N^{1.5}$ is contained in $\Lambda_{p}(\sqrt{0.4 \zeta \mu^2/N})$ and thus, we can guarantee pairwise $\epsilon$-isometry for this set with $M=O\left( \epsilon^{-2}\log\left( N \epsilon^{-1} (1-\tau \chi)^{-1} \zeta^{-1} \alpha^{-1} \right) \right)$ measurements.

Combining the isometries in the regimes $|\omega_{1}-\omega_{2}| \leq \mu/N^{1.5}$ and $|\omega_{1}-\omega_{2}| \geq \mu/N^{1.5}$ completes the proof of Theorem \ref{thm:singleSine}.

\negspace
\section{Extending tangent plane isometry}
\label{appendix:extendTgtPlane}
We first derive a bound on $\Vert \be \Vert$. Applying the triangle inequality to $\be$, we obtain $\Vert \be \Vert \leq |g_{1}| \Vert \be_{1} \Vert + |g_{2}| \Vert \be_{2} \Vert$. Since the quantity we wish to bound $\Vert \bA \bX(\bomega) \bg \Vert/\Vert \bX(\bomega) \bg \Vert$ does not depend on $\Vert \bg \Vert$, we can, without loss of generality, restrict attention to $\Vert \bg \Vert = 1$. Thus, we have $|g_{i}| \leq 1$. We use the mean value theorem to obtain bounds on $\Vert \be_{i} \Vert~i=1,2$ (the mean value theorem relates $\be_{i}$ to $d^{2}\bx(\omega_{i}')/d\omega^{2}$ for some $\omega_{i}' \in \left[\omega_{1},\omega_{2}\right]$) and ultimately get $\Vert \be \Vert \leq N^{2}\Delta^{2}/(4 \sqrt{2})$.

In order to obtain a lower bound for $\Vert \bv \Vert$, we rewrite $\bv$ as
\begin{equation}
\mathbf{v}=\mathbf{T}(q)
\left[\begin{array}{cc} \sqrt{2}&0\\
0&\frac{\Delta}{\sqrt{2}\tau}
\end{array}\right]
\left[\begin{array}{cc} \frac{1}{\sqrt{2}}&\frac{1}{\sqrt{2}}\\
\frac{-1}{\sqrt{2}}&\frac{1}{\sqrt{2}}
\end{array}\right]\left[\begin{array}{c} g_1\\ g_2\end{array}\right].\label{eq:small:pairwise:svd}
\end{equation}
We now recall that the minimum singular value of the product of two matrices is at least as large as the product of their minimum singular values. The minimum singular value of $\mathbf{T}(q)$ is $\sigma_{\mathrm{tangent}}=\sqrt{1-\tau\chi}$ and the corresponding value for the other two matrices are $|\Delta|/\sqrt{2}\tau$ and $1$ respectively. Thus, the minimum singular value of the product of the three matrices is greater than $\sqrt{1-\tau\chi} \times \frac{|\Delta|}{\sqrt{2}\tau}$. Since $\Vert \bg \Vert = 1$, we immediately get the desired bound on $\Vert \bv \Vert$.

\negspace
\section{Smallest singular value for well-separated frequencies}
\label{appendix:singValWellSep}
We wish to obtain a lower bound for the smallest singular value $\sigma_{\textrm{signal}}^{2}$ of the matrix $\left[\bX(\omega_{1})~\bX(\omega_{2})\right]$ when the frequencies satisfy $|\omega_{1}-\omega_{2}| > \mu/N^{1.5}$. First, we note that this is equivalent to upper-bounding $|H(\omega_{1}-\omega_{2})|$ since $\sigma_{\textrm{signal}}^{2} = 1-|H(\omega_{1}-\omega_{2})|$. Since $|H(\omega)|$ is not necessarily monotonic (imagine that $|h_{n}|^2$ is the Hamming window; $|H(\omega)|$, being the magnitude of the Fourier transform of $|h_{n}|^{2}$, has sidelobes), it is not true in general that that the maximum of $|H(\omega)|, |\omega| > \mu/N^{1.5}$ occurs at $\omega = \mu/N^{1.5}$. However, we now make two observations that allow us to analyze the behavior of $|H(\omega)|$ only at the minimum separation $\mu/N^{1.5}$. 

First, if there were no restrictions on the frequencies $(\omega_{1},\omega_{2})$, $|H(\omega_{1}-\omega_{2})|$ has a maximum ($=1$) when $\omega_{1}=\omega_{2}$. Second, because (i) the set of the frequencies we are excluding $|\omega_{1} - \omega_{2}| < \mu/N^{1.5}$ is very small (it is smaller than $\pi/(2N)$ for large enough $N$) and (ii) we restrict ourselves to sequences $\{h_n\}$ such that $|H(\omega)|$ is monotone in $(0,\pi/(2N))$, the maximum of $|H(\omega_{1}-\omega_{2})|, \pi/(2N)>|\omega_{1}-\omega_{2}| > \mu/N^{1.5}$ is guaranteed to occur when $\omega_{2} = \omega_{1} \pm \mu/N^{1.5}$.

For small values of $|\omega_{1}-\omega_{2}|$, we can expand $|H(\omega)|^2$ around $\omega = 0$ to get
\begin{equation}
|H(\omega_1 - \omega_2)|^2 = 1 - \zeta N^2\left(\omega_1-\omega_2\right)^2\pm
O\left(N^4\left(\omega_1-\omega_2\right)^4\right),
\end{equation}
where 
$\zeta=-({N^{-2}}/{2!})\left.{d^2 |H(\omega)|^2}/{d\omega^2} \right|_{\omega=0}$. For $|\omega_{1}-\omega_{2}| = \mu/N^{1.5}$, we have 
\begin{equation}
\left|H\left({\mu}/{N^{1.5}}\right)\right|^2 = 1 - \zeta \mu^2/N \pm O\left(1/N^2\right).
\end{equation}
we see that $|H(\mu/N^{1.5})|$ approaches $1$ with increasing $N$. Since we assume that all side-lobes are smaller than $D<1$, there exists some $N$ beyond which the maximum of $|H(\omega_{1}-\omega_{2})|, |\omega_{1}-\omega_{2}| >\pi/(2N)$ is guaranteed to be smaller than $|H(\mu/N^{1.5})|$. Therefore, for sufficiently large $N$, the maximum of $|H(\omega_1-\omega_2)|$ for all $|\omega_1-\omega_2|>\mu/N^{1.5}$ occurs at $|\omega_1-\omega_2|=\mu/N^{1.5}$. Plugging the expression for $|H(\omega_1-\omega_2)|$ in $\sigma_{\textrm{signal}}^{2}$ for this frequency separation, we have that, 
\begin{equation}
N\sigma_{\textrm{signal}}^{2} \geq 0.5\zeta \mu^2 \pm O \left(1/N\right). \label{eq:appendix:bound_sing}
\end{equation}
To arrive at the above expression we have used the following: $|H(\omega)|\leq 1 ~\forall \omega$ (since $\sum|h_n|^2 = 1$). This gives us 
\begin{equation}
1- |H(\omega)| \geq ({1+|H(\omega)|}) (1-|H(\omega)|)/{2}~\forall\omega. 
\end{equation}
Therefore, $\sigma_{\textrm{signal}}^{2}\geq (1-|H(\mu/N^{1.5})|^2)/2$, which yields (\ref{eq:appendix:bound_sing}). 
The first term in (\ref{eq:appendix:bound_sing}) given by $0.5\zeta \mu^{2}$, is a constant and the second term decays to zero (as $1/N$). Therefore, for large values of $N$, the second term is much smaller than the first and $\sigma_{\textrm{signal}}^{2}$ is bounded away from zero. In particular, for large enough $N$ (how large it needs to be depends on $\mu$ and the behavior of $|H(\omega)|^2$ at $\omega = 0$), we have $\sigma_{\textrm{signal}} > \sqrt{0.4 \zeta \mu^2/N}$.

\bibliographystyle{IEEEtran}
\bibliography{references}

\begin{thebibliography}{10}
\providecommand{\url}[1]{#1}
\csname url@samestyle\endcsname
\providecommand{\newblock}{\relax}
\providecommand{\bibinfo}[2]{#2}
\providecommand{\BIBentrySTDinterwordspacing}{\spaceskip=0pt\relax}
\providecommand{\BIBentryALTinterwordstretchfactor}{4}
\providecommand{\BIBentryALTinterwordspacing}{\spaceskip=\fontdimen2\font plus
\BIBentryALTinterwordstretchfactor\fontdimen3\font minus
  \fontdimen4\font\relax}
\providecommand{\BIBforeignlanguage}[2]{{%
\expandafter\ifx\csname l@#1\endcsname\relax
\typeout{** WARNING: IEEEtran.bst: No hyphenation pattern has been}%
\typeout{** loaded for the language `#1'. Using the pattern for}%
\typeout{** the default language instead.}%
\else
\language=\csname l@#1\endcsname
\fi
#2}}
\providecommand{\BIBdecl}{\relax}
\BIBdecl

\bibitem{yuejie_chi_sensitivity_Basis_Mismatch_2011}
Y.~Chi, L.~Scharf, A.~Pezeshki, and A.~Calderbank, ``Sensitivity to basis
  mismatch in compressed sensing,'' \emph{Signal Processing, IEEE Transactions
  on}, vol.~59, no.~5, pp. 2182--2195, 2011.

\bibitem{candes2}
E.~Candes and T.~Tao, ``Near-optimal signal recovery from random projections:
  Universal encoding strategies?'' \emph{Information Theory, IEEE Transactions
  on}, vol.~52, no.~12, pp. 5406 --5425, dec. 2006.

\bibitem{donoho}
D.~Donoho, ``Compressed sensing,'' \emph{Information Theory, IEEE Transactions
  on}, vol.~52, no.~4, pp. 1289 --1306, april 2006.

\bibitem{baraniuk_simple_RIP_2008}
R.~Baraniuk, M.~Davenport, R.~{DeVore}, and M.~Wakin, ``A simple proof of the
  restricted isometry property for random matrices,'' \emph{Constructive
  Approximation}, vol.~28, no.~3, p. 253–263, 2008.

\bibitem{achlioptas_database-friendly_2001}
D.~Achlioptas, ``Database-friendly random projections,'' ser. {PODS} '01.\hskip
  1em plus 0.5em minus 0.4em\relax New York, {NY}, {USA}: {ACM}, 2001, p.
  274–281.

\bibitem{baraniuk_random_2009}
R.~Baraniuk and M.~Wakin, ``Random projections of smooth manifolds,''
  \emph{Foundations of Computational Mathematics}, vol.~9, no.~1, p. 51–77,
  2009.

\bibitem{wakin_manifold_estimation_2010}
M.~Wakin, ``Manifold-based signal recovery and parameter estimation from
  compressive measurements,'' \emph{Arxiv preprint arxiv:1002.1247}, 2010.

\bibitem{Duarte_2012_SpectralCS}
M.~F. Duarte and R.~G. Baraniuk, ``Spectral compressive sensing,''
  \emph{Applied and Computational Harmonic Analysis}, 2012.

\bibitem{fannjiang:superresolution:compressive}
A.~Fannjiang and W.~Liao, ``Coherence pattern-guided compressive sensing with
  unresolved grids,'' \emph{SIAM Journal on Imaging Sciences}, vol.~5, no.~1,
  pp. 179--202, 2012.

\bibitem{asilomar:CRLB:frequency}
D.~Ramasamy, S.~Venkateswaran, and U.~Madhow, ``Compressive estimation in
  {AWGN}: {G}eneral observations and a case study,'' in \emph{Signals, Systems
  and Computers (ASILOMAR), 2012 Conference Record of the Forty Sixth Asilomar
  Conference on}, 2012, pp. 953--957.

\bibitem{allerton:tracking}
------, ``Compressive tracking with 1000-element arrays: {A} framework for
  multi-{G}bps mm wave cellular downlinks,'' in \emph{Communication, Control,
  and Computing (Allerton), 2012 50th Annual Allerton Conference on}, 2012, pp.
  690--697.

\bibitem{Recht_2013_with_Noise_no_CS}
G.~Tang, B.~N. Bhaskar, and B.~Recht, ``Near {M}inimax {L}ine {S}pectral
  {E}stimation,'' \emph{CoRR}, vol. abs/1303.4348, 2013.

\bibitem{Recht_2012_denoising}
B.~N. Bhaskar, G.~Tang, and B.~Recht, ``Atomic norm denoising with applications
  to line spectral estimation,'' \emph{CoRR}, vol. abs/1204.0562, 2012.

\bibitem{2012_Recht_CS_Off_the_Grid}
G.~Tang, B.~N. Bhaskar, P.~Shah, and B.~Recht, ``{C}ompressed {S}ensing off the
  {G}rid,'' \emph{CoRR}, vol. abs/1207.6053, 2012.

\bibitem{Eldar_Noise_folding_2011}
E.~Arias-Castro and Y.~Eldar, ``{N}oise {F}olding in {C}ompressed {S}ensing,''
  \emph{Signal Processing Letters, IEEE}, vol.~18, no.~8, pp. 478--481, 2011.

\bibitem{nowak:CompressiveDetection}
J.~Haupt and R.~Nowak, ``Compressive sampling for signal detection,'' in
  \emph{Acoustics, Speech and Signal Processing, 2007. ICASSP 2007. IEEE
  International Conference on}, vol.~3, april 2007, pp. III--1509 --III--1512.

\bibitem{ITA2012}
D.~Ramasamy, S.~Venkateswaran, and U.~Madhow, ``Compressive adaptation of large
  steerable arrays,'' in \emph{Information Theory and Applications Workshop
  (ITA), 2012}, 2012, pp. 234--239.

\bibitem{maravic_vetterli_FRI_2005}
I.~Maravic and M.~Vetterli, ``Sampling and reconstruction of signals with
  finite rate of innovation in the presence of noise,'' \emph{Signal
  Processing, IEEE Transactions on}, vol.~53, no.~8, pp. 2788--2805, 2005.

\bibitem{Eldar_FRI_sampling_CRB_2012}
Z.~Ben-Haim, T.~Michaeli, and Y.~Eldar, ``{P}erformance {B}ounds and {D}esign
  {C}riteria for {E}stimating {F}inite {R}ate of {I}nnovation {S}ignals,''
  \emph{Information Theory, IEEE Transactions on}, vol.~58, no.~8, pp.
  4993--5015, 2012.

\bibitem{eldar_2012Xampling}
M.~Mishali and Y.~C. Eldar, ``{X}ampling: {C}ompressed {S}ensing for {A}nalog
  {S}ignals,'' in \emph{{C}ompressed {S}ensing: {T}heory and {A}pplications},
  Y.~C. Eldar and G.~Kutyniok, Eds.\hskip 1em plus 0.5em minus 0.4em\relax
  Cambridge University Press, 2012.

\bibitem{van2013detection}
H.~L. Van~Trees, K.~L. Bell, and Z.~Tian, \emph{{D}etection, {E}stimation, and
  {M}odulation Theory}.\hskip 1em plus 0.5em minus 0.4em\relax John Wiley \&
  Sons, 2013.

\bibitem{vershynin_introduction_2010}
R.~Vershynin, ``Introduction to the non-asymptotic analysis of random
  matrices,'' {arXiv} e-print, Nov. 2010, chapter 5 of: Compressed Sensing,
  Theory and Applications. Edited by Y. Eldar and G. Kutyniok. Cambridge
  University Press, 2012.

\bibitem{book::2007::vantrees::bayesian}
H.~Van~Trees and K.~Bell, ``Bayesian bounds for parameter estimation and
  nonlinear filtering and tracking,'' 2007.

\bibitem{Bell::EZZB::1997}
K.~Bell, Y.~Steinberg, Y.~Ephraim, and H.~Van~Trees, ``Extended {Z}iv-{Z}akai
  lower bound for vector parameter estimation,'' \emph{Information Theory, IEEE
  Transactions on}, vol.~43, no.~2, pp. 624 --637, mar 1997.

\bibitem{rife:boorstyn:CRB:freq}
D.~Rife and R.~Boorstyn, ``Single tone parameter estimation from discrete-time
  observations,'' \emph{Information Theory, IEEE Transactions on}, vol.~20,
  no.~5, pp. 591--598, 1974.

\bibitem{Basu::ZZB::Periodic::2000}
S.~Basu and Y.~Bresler, ``A global lower bound on parameter estimation error
  with periodic distortion functions,'' \emph{Information Theory, IEEE
  Transactions on}, vol.~46, no.~3, pp. 1145--1150, 2000.

\bibitem{candes:superresolution}
E.~Candes and C.~Fernandez-Granda, ``Towards a mathematical theory of
  super-resolution,'' \emph{CoRR}, vol. abs/1203.5871, 2012.

\bibitem{Weiss_1994}
A.~Weiss and B.~Friedlander, ``Preprocessing for direction finding with minimal
  variance degradation,'' \emph{Signal Processing, IEEE Transactions on},
  vol.~42, no.~6, pp. 1478--1485, 1994.

\bibitem{ferreira_super-resolution_1999}
P.~Ferreira, ``Super-resolution, the recovery of missing samples and
  vandermonde matrices on the unit circle,'' in \emph{Proceedings of the
  Workshop on Sampling Theory and Applications, Loen, Norway}, 1999.

\end{thebibliography}

\end{document}